\documentclass[useAMS,usenatbib,times]{mn2e}
\usepackage{mathpazo}

\usepackage{graphicx}
\usepackage[usenames]{color}
\definecolor{Blue}{rgb}{0,0.08,0.65}
\definecolor{Red}{rgb}{0.65,0.08,0.05}
\definecolor{Green}{rgb}{0.15,0.45,0.25}
\definecolor{Green2}{rgb}{0.15,0.85,0.15}
\usepackage{hyperref}
\hypersetup{dvips, colorlinks=true, linkcolor=black, citecolor=black, filecolor=black, urlcolor=blue}

\newcommand{\nut}{\mbox{{\sc \small Nut}}}

\def\gtrsim{\lower.5ex\hbox{$\; \buildrel > \over \sim \;$}}
\def\lesssim{\lower.5ex\hbox{$\; \buildrel < \over \sim \;$}}

\title[Connecting environment of dark matter halos to their spin]{Connecting  the cosmic web to the spin of dark halos:\\
implications for galaxy formation }

\author[S. Codis, C. Pichon, J. Devriendt, A. Slyz, D. Pogosyan, Y. Dubois, T. Sousbie ]{\parbox[t]{\textwidth}{
Sandrine Codis$^{1}$, Christophe Pichon$^{1,2}$, Julien Devriendt$^{2}$,  Adrianne Slyz$^{2}$, 
\\
Dmitry Pogosyan$^{3}$, 
Yohan Dubois$^{1}$,  and 
Thierry Sousbie$^{1}$  }\\
$^{1}$ Institut d'Astrophysique de Paris, 98 bis boulevard Arago, 75014 Paris, France\\
$^{2}$ Astrophysics  sub department, University of Oxford, Keble Road, Oxford OX1 3RH, UK. \\
 $^{3}$Department of Physics, University of Alberta, 11322-89 Avenue, Edmonton, Alberta, T6G 2G7, Canada.
}

\begin{document}

\pagerange{\pageref{firstpage}--\pageref{lastpage}} \pubyear{2010}
\maketitle
\label{firstpage}

\begin{abstract}
We investigate the alignment of the spin of  dark matter halos relative (i) to the surrounding large-scale filamentary structure, and (ii) to the tidal tensor eigenvectors 
 using the Horizon 4$\pi$ dark matter simulation which resolves over 43 million dark matter halos at redshift zero. 
 We detect a clear mass transition: the spin of dark matter halos above a critical mass $M^s_{0} \approx 5(\pm 1)\cdot 10^{12} M_\odot$ tends to be perpendicular to the closest large scale filament (with an excess probability up to 12\%), and aligned with the intermediate
 axis of the tidal tensor (with an excess probability of up to 40\%), whereas  the spin of low-mass halos is more likely to be aligned with the closest filament (with an excess probability up to 15\%).  Furthermore, this critical mass is  redshift-dependent, scaling as 
 $M^s_{\rm crit} (z)\approx M^s_{0}\cdot (1+z)^{-\gamma_s}$ with $\gamma_s=2.5 \pm 0.2$. A similar fit for the redshift evolution of the tidal tensor transition mass yields $M^t_{0} \approx8 (\pm 2) \cdot 10^{12} M_\odot$ and $\gamma_t=3 \pm 0.3$. This critical mass also varies weakly with the scale defining filaments.

   We propose an interpretation of this signal in terms of large-scale cosmic flows. In this picture,
  most low-mass halos are formed through the winding of flows embedded in misaligned walls; hence they acquire a spin parallel to the axis of the resulting filaments
forming at the intersection of these walls. 
  On the other hand, more massive halos are typically the products of  later mergers along such filaments,  and thus they acquire a spin perpendicular to this direction when their orbital angular momentum 
  is converted into spin.
We show that this scenario is  consistent with both the measured excess probabilities of alignment w.r.t. the eigen-directions of the tidal tensor, and 
halo merger histories. On a more qualitative level, it also seems compatible with 3D visualization of the structure of the cosmic web as traced by ``smoothed'' dark matter simulations or gas tracer particles.
Finally, it provides extra support to the disc forming paradigm presented by \cite{pichonetal11} as it extends it  by characterizing the geometry of secondary infall at high redshift.

\end{abstract}

\begin{keywords}
large-scale structure -- galaxies: halos -- galaxies: formation -- method: numerical.
\end{keywords}

\section{Introduction}
\label{sec:intro}

Over the past decades, numerical simulations and large redshift surveys have highlighted the large-scale structure 
 of our Universe (LSS), a cosmic web formed by voids, sheets, elongated filaments and clusters at their nodes \citep{bkp96}. This structure is believed to be the result of the linear growth of primordial gaussian fluctuations in a nearly homogeneous density field followed by the non-linear collapse of overdense regions into dark matter halos which then accrete mass and merge as described by the hierarchical model. 
 The current paradigm of galaxy formation states that collapsing protogalaxies acquire their spin (ie their angular momentum) by tidal torquing because of a misalignment between their inertia tensor and the local gravitational tidal tensor at the time of maximum expansion;
this is the basis of the so-called  Tidal Torque Theory \citep[TTT hereafter,][for a recent review]{b0,b1,b3,b2,b31,b32,schafer09}. 
 According to this theory, the spin direction should initially  be correlated with the principal axes of the local tidal tensor,  defined as the traceless part of the Hessian matrix of the gravitational potential field. 
One therefore expects to detect correlations between actual galactic angular momenta and large-scale structures if non-linear processes have not modified their direction.

For many years, both observers and theorists have thus endeavoured to detect these correlations in real surveys and cosmological N-body simulations. Nonetheless, the results remain in part contradictory because of the lack in resolution together with the difficulty of properly defining large-scale filamentary structures.

For instance, using N-body simulations \cite{b12}, \cite{b7} and \cite{b18} found that halo spins are preferentially oriented perpendicularly to the filaments whatever their mass, with \citet{b23} measuring a random distribution of the spin in the plane perpendicular to the filaments, while \citet{b33} claimed an alignment between spin and filament.
 More
recently a consensus seemed to have emerged when several works \citep{b24,b10,b14,b16} reported that large-scale structure - filaments and sheets - influenced the direction of the  angular momenta of  dark matter halos in a way originally predicted by \citet{b8} and \cite{b9}. These studies pointed towards the detection of a mass-dependence orientation of the spin, arguing for the first time that the spin of high-mass halos tends to lie perpendicular to their host filament, whereas low-mass halos have a spin preferentially aligned with it.
Nevertheless, the detected correlation remains weak and noisy and no full explanation for these findings were  highlighted except e.g. \cite{b24} who suggested that the spin direction of the cluster and group mass halos (as opposed to galaxy mass halos) could come from mergers along the filaments.

However, \cite{hahn10} repeated this measurement in a cosmological hydrodynamical re-simulation of one large-scale cosmic filament and found a different result: the spin of high-mass halos is
 aligned with the filament;   the spin of low-mass halos is along the intermediate eigen-direction of the tidal tensor in low-density regions and along the third eigen-direction (i.e neither the intermediate direction nor the filament's) in higher density regions at higher redshift (z=1); finally no signal exists for high-density region at low redshift.

Beyond numerical simulations, \citet{b13,trujillo06,navarro04,flin90,flin86,godlowski10} found correlations in observations between the rotational axis of galaxies and the surrounding large-scale structures (e.g voids and local tidal shears) unlike \cite{b34} who did not find any correlation. To be more specific, \cite{flin86,flin90} first discovered that the spin of galaxies was not isotropically oriented w.r.t the Local Supercluster plane but more likely to be aligned with it; \cite{navarro04} confirmed this observation; \cite{trujillo06} found that in the SDSS and 2dFGRS, the rotational axis of the spiral galaxies located in the walls surrounding voids lie preferentially in the plane of these voids; \cite{b13} analyzed the galaxies of the 2Mass Redshift Survey and found correlations between their spin and the local tidal tensor and \cite{godlowski10} focused on the galaxy groups in the Local Supercluster and found correlations in their orientation suggesting that the two brightest galaxies and then the galaxy groups were hierarchically formed in the same filament with their major vector aligned with this host filament . The results of \cite{navarro04,trujillo06} support the predictions of TTT \citep{b9} namely that, assuming that the inertia and tidal tensors are uncorrelated, galaxies's spin should be preferentially aligned with the intermediate eigen-direction of the tidal tensor (in particular in the plane of the voids). Recently, however, \cite{slosar09} claimed that in contrast to previous studies, they found no departure from randomness in the SDSS while studying the orientation of the galaxy spin with regards to the voids in which they are located. The methods used in the latter study has been improved by \cite{varela11} who used the SDSS (DR 7) and morphological classification from the Galaxy Zoo Project and found that galaxy disks are more likely to lie in the plane of voids i.e. their spin tend to be perpendicular to the void they are located in, which seems in disagreement with  \citet{b13,trujillo06,navarro04}.

 In short, even if one can claim that a trend is slowly emerging, quantitative evidence of spin alignment with  the filaments and tidal tensor eigen-directions remains at this stage weak and somewhat inconclusive.
Hence, in this paper, we propose to revisit the issue and quantify the alignment between the spin of dark matter halos and the filamentary pattern in which they are embedded (together with the alignment between the spin and the tidal tensor principal axes) using a very efficient topological tool, the skeleton  \citep{b7}. This tool provides a robust and mathematically well defined reconstruction of the cosmic web filaments. 
We apply it to the Horizon 4$\pi$ simulation \citep{b4}, a 2 h$^{-1}$Gpc on a side cubic volume of the universe containing over 67 billion dark matter particles which provides an unprecedented catalog of 43 million dark matter halos with masses $> 2 \cdot 10^{11}$M$_\odot$. We then interpret our results in the framework of the dynamics  of large-scale cosmic flows. 

Section~\ref{sec:fil-spin} briefly presents the Horizon 4$\pi$ simulation  and the topological tool implemented to identify the loci and orientation of filaments. 
It then reports the correlations detected between the orientation of the spin of dark matter halos and filaments and its redshift evolution.
Section~\ref{sec:discussion}  is devoted to the physical processes which induce these correlations. It also illustrates them using dark matter halo merging histories, smoothed dark matter simulations 
and hydrodynamical simulations. 
Section~\ref{sec:conclusion} provides conclusions and discusses prospects for our understanding of  galaxy formation within its cosmic environment.
 Appendix~\ref{sec:shear} gives the correlations measured between spin direction and tidal eigen-directions, which are in agreement with the cosmic dynamics arguments of Section~\ref{sec:fil-spin}. 
 Appendix~\ref{sec:ap-hydro} presents a visual quantitative estimation of the spin of the circum-galactic medium. 
 Appendix~\ref{sec:checks} sums up all the tests we have performed to assess the robustness of the  correlations presented in Section~\ref{sec:fil-spin}.
 Finally, Appendix~\ref{sec:masses} presents  the dependence of the transition masses with the smoothing length and 
 the non-linear mass as a function redshift. 
 
\section{Spin-filament correlations}
\label{sec:fil-spin}
Let us first account for the robust correlation between the dark matter halo's spin and the orientation of the filaments of the cosmic web.
\subsection{Virtual data sets}
\label{sec:simus}
This study uses the Horizon 4$\pi$ N-body simulation \citep{b4} which contains $4096^3$ dark matter particles distributed in a 2 $h^{-1}$Gpc periodic box to investigate 
the spin alignment of dark matter halos relative to their large-scale structure environment. This simulation is characterized by the following $\Lambda$CDM cosmology: $\Omega_{\rm m}=0.24 $, $\Omega_{\Lambda}=0.76$, $n=0.958$, $H_0=73 $ km$\cdot s^{-1} \cdot $Mpc$^{-1}$ and $\sigma _8=0.77$ within one standard deviation of WMAP3 results \citep{Spergeletal03}. 
These initial conditions were evolved non-linearly down to redshift zero using the AMR code RAMSES \citep{b26}, on a $4096^3$ grid. The motion of the particles was followed with a multi grid Particle-Mesh Poisson solver using a Cloud-In-Cell interpolation algorithm to assign these particles to the grid (the refinement strategy of  40 particles as a threshold for refinement allowed us to reach a constant physical resolution of 10 kpc, see the above mentioned two references).

\begin{figure}
 \includegraphics[width=0.98\columnwidth]{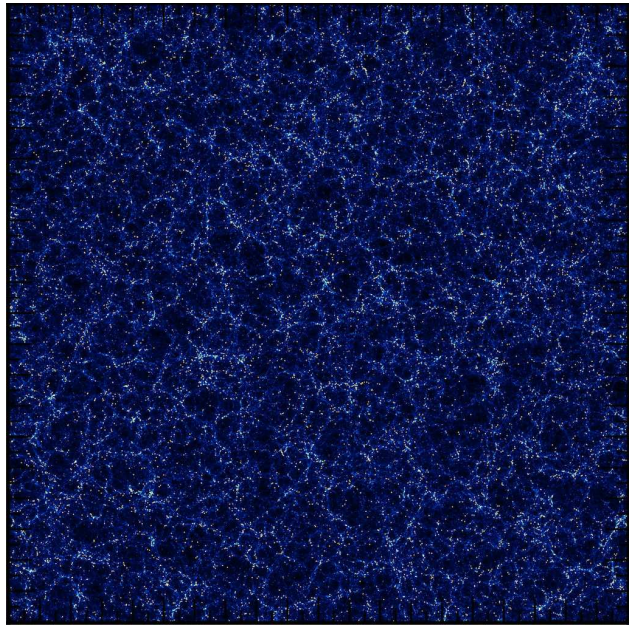}
 \caption{An 80 Mpc/$h$ slice of the   Horizon 4$\pi$ simulation at redshift zero.  The box-size is 2Gpc$/h$ across. On top of  the dark matter log-density (colour coded in levels of blue, from dark to light),
 all  halos in that slice  whose mass is larger than $3\cdot 10^{13} M_\odot$ are shown as yellow dots. As expected, these halos fall on top of the filamentary structure of the cosmic 
 web.
   \label{fig:peaks}}
\end{figure}

The Friend-of-Friend Algorithm \citep[hereafter FOF,][]{b27} was used  over $18^3$ overlapping subsets of the simulation with a linking length of  0.2 times the mean inter-particular distance  to define dark matter halos. 
In the present work, we only consider halos with more than 40 particles, which corresponds to a minimum halo mass of $3 \cdot 10^{11} M_{\odot}$ (the particle mass is $7.7\cdot 10^{9}M_{\odot}$). The mass dynamical range of this simulation spans about 5 decades.
Overall, $43$ million halos were detected at redshift zero (see fig.~\ref{fig:peaks}). 
This simulation was complemented by smaller ($1024^3$ particles, boxsize 200 $h^{-1}$Mpc and several $256^3$ particles, boxsize 50 $h^{-1}$Mpc leading to a particle mass of $6.2 \cdot 10^{8}M_{\odot}$) 
dark matter only simulations to address resolution issues (see Appendix~\ref{sec:checks}) and interpret the redshift evolution of the signal (see Section~\ref{sec:redshift}).
%
\begin{figure}
 \includegraphics[width=0.98\columnwidth]{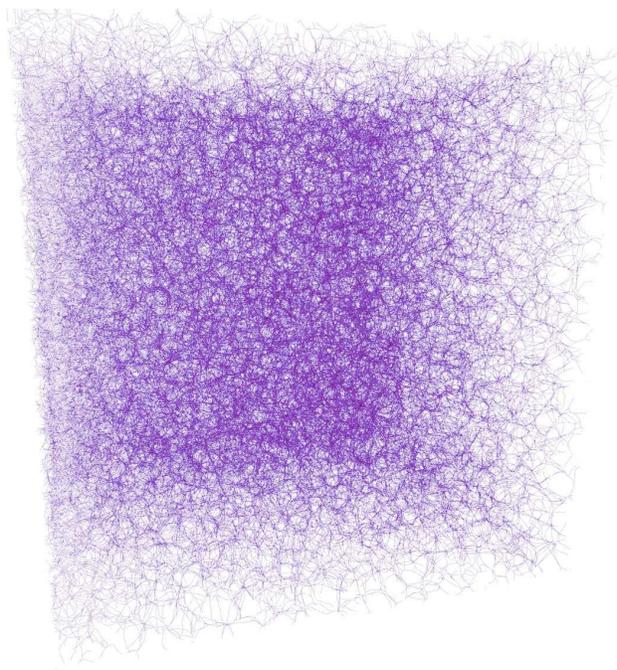}
 \caption{A 3D view of the skeleton of  the Horizon 4$\pi$ simulation measured from the dark matter distribution.  The size of the box is $2~Gpc/h$.
 This paper analyses the relative orientation of the spin of the 43 million dark halos
 relative to this cosmic web.
  \label{fig:skel}}
\end{figure}
%

Several topological techniques \citep{b29,b30,b5}  
have been proposed to identify the complex cosmic network made of large voids, sheet-like structures and elongated filaments. These techniques rely on giving a mathematical definition (and a detection algorithm) of the filamentary pattern that our eye easily detects in the simulations. One recent criterion to classify structures as clusters, filaments, sheets or voids, uses the number of positive eigenvalues of the Hessian matrix of the density or potential fields \citep{b14,b10,b19}. Another interesting approach was followed by \citet{b20,b21} who used Watershed transforms to identify voids, filaments and walls. More recently \cite{gonzalez10} introduced a method which relies on positions and masses of dark matter halos, and \cite{bond10}  used Hessian eigen-directions to detect filaments.
In this paper, we use the (global) 3D skeleton introduced by \citet{b7}. The underlying  algorithm is based on Morse theory and defines the skeleton as the set of critical lines joining the maxima of the density field through saddle points following the gradient. 
In practice \citet{b7} 
 define the peak and void patches of the density field as the set of points converging to a specific local maximum/minimum while following the field lines in the direction/opposite direction of the gradient. The skeleton is then the set of intersection of the void patches i.e. the subset of critical lines connecting the saddle points and the local maxima of a density field and parallel to the gradient of the field.

\label{sec:method}
For this work, the $\sim$70 billion particles of the Horizon 4$\pi$ were sampled on a $2048^3$ cartesian grid and 
the  density field was smoothed over 5 sigmas using {\tt mpsmooth} \citep{b28}, corresponding to  a scale of 5 Mpc$/h$ and a mass of 
$1.9 \times 10^{14} M_\odot$. Hence we are focusing on the large-scale structure of the cosmic web.  The corresponding cube was then divided into  $6^3$ overlapping sub-cubes  \cite[with a buffer zone of 100 voxels in each direction, large enough to 
 cover the largest peak patches of the simulation, see][]{b7}, and  the skeleton was computed for each of these sub-cubes. It was then reconnected across
the entire simulation volume to produce a catalog of segments which locally defines the direction of the skeleton. This skeleton is shown in fig.~\ref{fig:skel}. Note that this skeleton (i.e what we will call filaments in the rest of this paper) depends on the choice we made for the smoothing length (5 Mpc$/h$).
Appendix~\ref{sec:smoothing} investigates the 
effect of probing  smaller smoothing scale  on other sets of simulation.

The hydrodynamical simulations used in this paper are described in Appendix~\ref{sec:ap-hydro}.

\subsection{Correlations between spin and filament axis}
 \label{sec:spin-fil}
In order to study the alignment between the spin of halos and the filamentary features of the cosmic web, we compute the skeleton of the LSS 
for the density field smoothed with  the  above quoted Gaussian scale $R=5~\mathrm{Mpc}/h$
which corresponds to $\sigma(R) = 0.66$. Thus we are considering the filaments
that are mildly non-linear large-scale structures at cluster scales.  

\begin{figure*}
 \includegraphics[width=0.98\columnwidth]{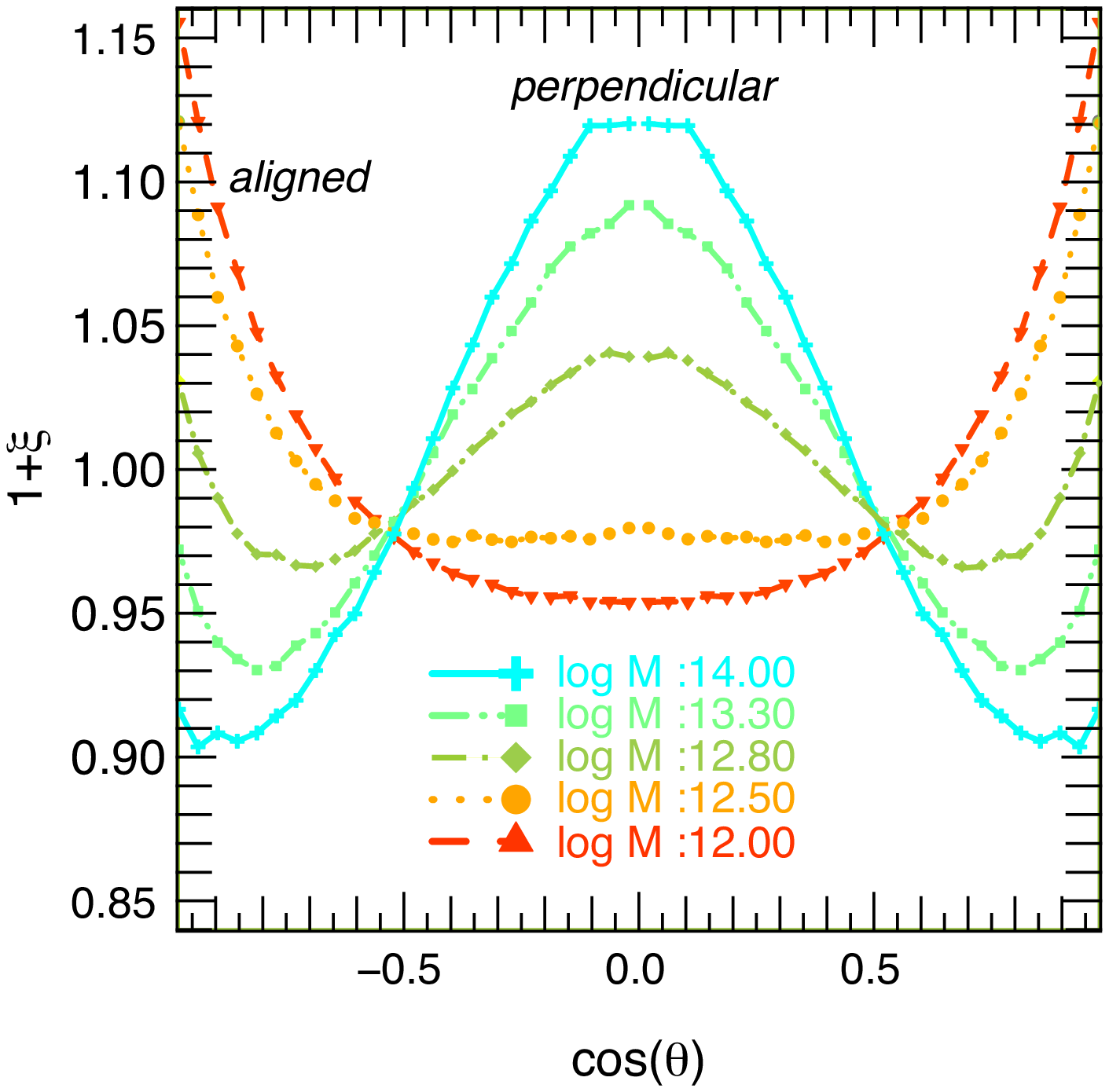}
 \hskip 0.4cm
 \includegraphics[width=1.025\columnwidth]{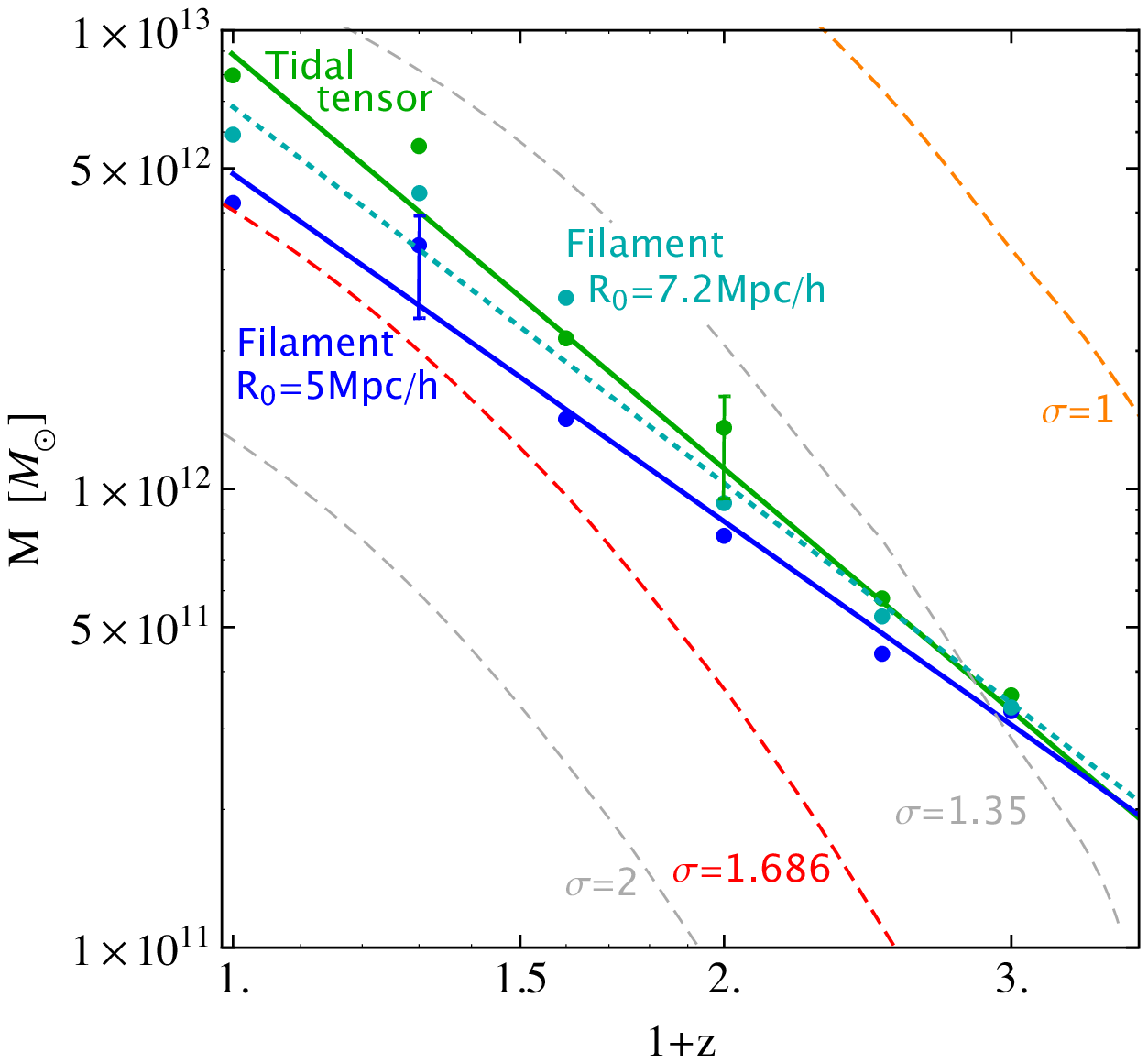}
 \caption{{\sl Left:} 
  excess probability of alignment  between the spin and the direction of the closest filament as measured from the 43 millions halos of  the Horizon 4$\pi$ simulation at redshift zero. 
  Different colors correspond to different mass bins from $ 10^{12}$({\sl  red}) to $ 10^{14}$ $M_\odot$ ({\sl blue})  as labeled.
   Thanks to the very large number of halos in each mass bin, the excess probability is quite well sampled and displays a clear departure from a uniform distribution.
  A transition  mass is detected at $M_{0}^{s}= M_{\rm crit}^{s}(z=0)\simeq 5 (\pm 1)\cdot 10^{12} M_{\odot}$: for halos with $M>M_{0}^{s}$,  the spin is more likely to be perpendicular to their host filament, whereas for halos with $M<M_{0}^{s}$, the spin tends to be aligned with the closest filament.  
    {\sl Right:} 
      redshift evolution of {\sl resp.} the filament transition mass (in {\sl blue})  and 
    the tidal tensor transition mass (in {\sl green}) derived from the 200 Mpc$/h$  $\Lambda$CDM simulations as discussed in the main text. The ({\sl cyan}) dotted line represents the spin-filament mass transition for a larger smoothing (7.2 Mpc/h).
   The  displayed error bar is estimated as a 1/3 of the bin mass.
     The dashed lines correspond to the non-linear masses (for a top-hat filter, see Appendix~\ref{sec:masstransition}) at different sigmas, in particular $\sigma=1$ ({\sl orange}) and $\sigma=1.686$  ({\sl red}).
   The redshift evolution of the transition masses are in qualitative agreement with that of $M_{\rm NL}(\sigma\lesssim 1.686)$ though they seem to remain close to power-laws throughout the explored range of redshift.
  \label{fig:mass}}
\end{figure*}

In this paper, the spin of a given halo is defined as the following sum over its particles i: $m_{p}\sum_{i}(\mathbf{r_{i}}-\bar\mathbf{r})\times(\mathbf{v_{i}}-\bar \mathbf{v})$ where $\bar \mathbf{r} $ is the center of mass of the FOF and $\bar \mathbf{v}$ its mean velocity.
We search for the five dark matter halos (regardless of their mass)
closest to  {\sl  each} filament segment (see Appendix~\ref{sec:checks} for 
alternative choices). We then  measure the angle between the angular momentum
of these halos and the direction of the filament segment 
and estimate the  probability distribution (PDF) of the absolute value of the
cosine of this angle; this PDF, $1+\xi$,  measures the excess probability of alignment between the halo spin and the direction of the filament  (note in particular that it is
normalized for $\cos \theta$ between 0 and 1; for aesthetic purpose only,  data are symmetrically plotted for $\cos \theta$ between $-1$ and 1;  Appendix~\ref{sec:checks} briefly discusses the associated biases). 
 The data is split by halo masses ranging from galactic to cluster masses 
 and is displayed in fig. \ref{fig:mass}, the main result of this paper.

\begin{figure}
 \includegraphics[width=\columnwidth]{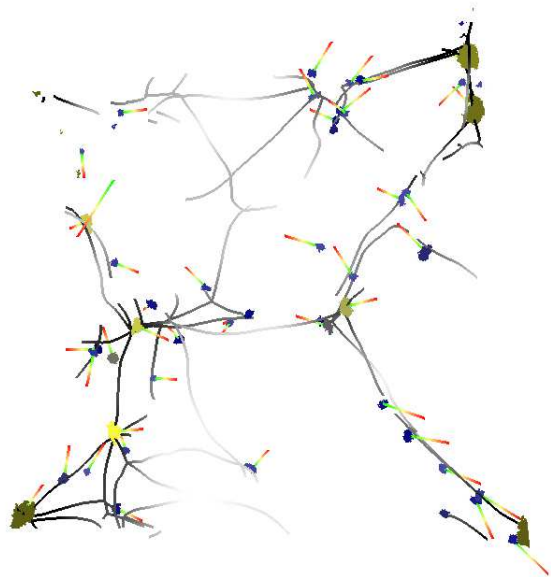}  \caption{ A very small subset of skeletons at different redshifts extracted from the Horizon 4$\pi$ simulation (see fig.~\ref{fig:skel}),
  together with unit vectors showing the orientation of the spin of the corresponding dark matter halo with a mass above the transition mass. The spin is indeed perpendicular to the filament for these massive halos.\label{fig:fig3d}
 }
\end{figure}

 A clear signal  is detected. The orientation of the halo spin depends
 on the local anisotropy of the cosmic web,{ \sl and} on the dark matter
 halo mass: the spin of dark matter halos is preferably perpendicular
 to their host filament at high mass (with an excess probability reaching
 12 \%), but turns into being aligned with the nearest filament direction
 at lower masses (with an excess probability of 15 \%). 
   This ``phase-transition" is found to occur 
   at $M^s_{\rm crit}(z=0)\simeq 4(\pm 1) \cdot10^{12} M_{\odot}$ where 
    $M^s_{\rm crit}$  is {\sl  defined } as the halo mass
   for which $\langle \cos\theta \rangle = 0.5$.
   Fig.~\ref{fig:fig3d} shows an example segment of the large-scale filamentary
   network together with orientation of spins of massive haloes that
   graphically demonstrates for them the effect of spin-filament
   anti-alignment. 
   Several sanity checks have been carried out to assess the robustness of this signal and are summed up in Appendix \ref{sec:checks}.

   These measurements of the spin-filament correlation trend 
   confirm the previous results obtained by \cite{b24,b10,b14,b16} with significantly improved statistics which allows us to quantify here the mass transition.

\subsection{Redshift dependence of transition mass}
\label{sec:redshift}
The redshift-dependence of the transition mass was then investigated on 
a set of smaller $\Lambda$CDM simulation ($1024^3$ particles 
in  $200$ $h^{-1}$Mpc periodic boxes and $256^3$ particles in
$50$ $h^{-1}$Mpc periodic boxes).
At high redshift we define the filamentary structure at the smoothing scale,
$R(z)$, chosen to maintain the same level of non-linearity as we had at
redshift zero for $R_{0}=5{\rm Mpc}/h$.   Thus, the smoothing scale $R(z)$
is obtained from the implicit condition
$
\sigma^2(R(z),z) =  \sigma^{2}(R_{0},0)
$, see Appendix~\ref{sec:masstransition}.
%
The halo spins continue to exhibit a transition from alignment with the nearest 
filament at low mass to anti-alignment at high mass. 
The critical mass for the transition, $M^s_{\rm crit}(z)$, is found to decrease
with redshift as a power-law of $z$. Namely, 
\begin{equation}
 M^s_{\rm crit}\approx M^s_{0} (1+z)^{-\gamma_s},\,  \, \gamma_s =2.5 \pm 0.2 \textrm{ , }\,\, M^s_0\simeq 5\cdot 10^{12} M_{\odot}. \label{eq:trendz}
\end{equation}
This result is presented in Fig.~\ref{fig:mass} over the studied range 
\mbox{$z=0-4$}.
Measuring the dependence of the $z=0$ transition mass
$M_{0}$ on scale $R_0$ (see Fig.~\ref{fig:smoothing}), we find some weak scaling, 
$M_0^s \propto R_0^{0.8}$. Note that this dependence significantly depends on  redshift.

The existence and redshift-dependence of the transition mass in 
 spin -- structure alignment is supported
by studying  the halo's spin direction relative 
to the orientation of the large-scale gravitational tidal tensor.
The details are given in Appendix~\ref{sec:shear}, where we find
that the more massive halos are preferably aligned with the intermediate principal
axis of the tidal tensor, while smaller halos show a positive alignment with
the minor axis (which near a filament is 
the direction in which the filament extends).
In this approach, our measurements give for the transition mass:
\begin{equation}
 M^t_{\rm crit}\approx M^t_{0} (1+z)^{-\gamma_t},\,  \, \gamma_t=3 \pm 0.3 \textrm{ , }\,\, M^t_0\simeq 8 \cdot 10^{12} M_{\odot}, \label{eq:trendz2}
\end{equation}
in good agreement with the skeleton results, equation (\ref{eq:trendz}).
The somewhat larger amplitude of $M_0^t$ w.r.t $M_{0}^{s}$ can be explained by noticing that
the tidal tensor associated with the gravitational potential 
smoothed on a scale $R_0$ effectively probes larger scales   
than the smoothed density field itself. 
Fig.~\ref{fig:mass} shows that
if we boost the smoothing scale used to define the skeleton to 
$R_0=7.2~\mathrm{Mpc}/h$,
$M_\mathrm{crit}^s$ and $M_\mathrm{crit}^t$ will match very closely (see also Appendix~\ref{sec:smoothing}).

In fig.~\ref{fig:mass}, we also compare $M_\mathrm{crit}(z)$ to
the redshift evolution of the mass scale $M_{\rm NL}(z)$ that corresponds
to the fixed $\sigma(R,z)$ (defined in Appendix~\ref{sec:masstransition}).
Several values of the variance are of interest to track.
One is 
 $\sigma(R,z)=1$, that formally defines the scale of non-linearity.
Another is $\sigma(R,z)=1.686$ that corresponds to the
characteristic mass scale, $M_\star$, of collapsed gravitationally bound halos
at redshift $z$ in the spherical top-hat model. 
Even though it is clear from fig.~\ref{fig:mass} 
that both transition masses $M_\mathrm{crit}^s$ and $M_\mathrm{crit}^t$
qualitatively match a non-linear mass evolution 
with  $\sigma \lesssim 1.686 $ at low redshifts, at high $z$ they
still follow a power-law behaviour, while $M_\mathrm{NL}$ steepens
as it probes the steepening power spectrum at ever shorter scales.
Thus, at high redshifts, the positive
alignment between the halo's spins and the nearby filaments 
extends to masses that are effectively higher,
in terms of  the corresponding characteristic non-linear mass.
Although we do not have the full quantitative explanation for this 
effect, it may be related to the fact that the filaments at high-z are
more pronounced due to a steeper power spectrum, and are correlated with 
the shear of the surrounding flow more robustly.  Note that the detection of halos and filaments at these redshifts may  be a concern for these intermediate resolution simulations.
Whilst we defer a detailed quantitative understanding of the redshift evolution of the mass transition,
the rest of the paper is devoted to {\sl explaining} the origin of these mass transitions.

\section{Spin induced by LSS dynamics}
\label{sec:discussion}

\begin{figure*}
   \includegraphics[width=1.5\columnwidth]{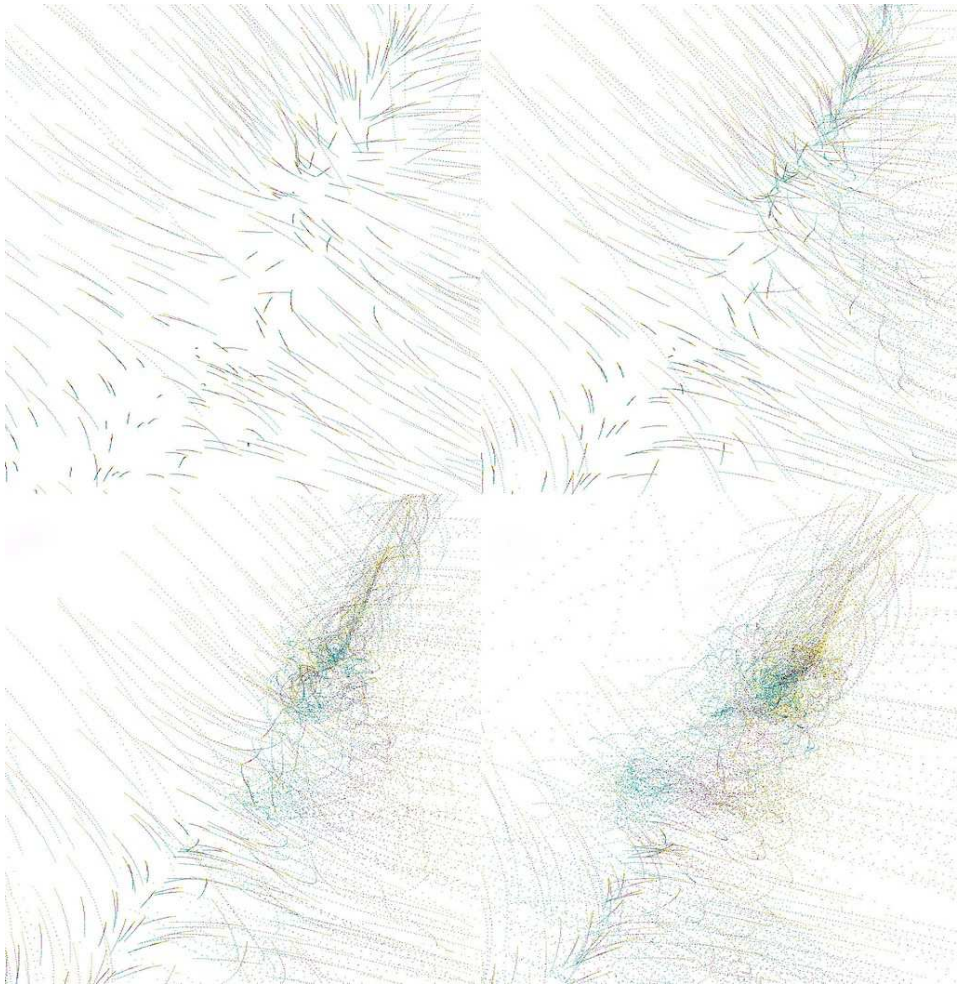}  	
\caption{Trajectory of dark matter particles, colour coded from {\sl yellow} (early) to {\sl blue} (late) via {\sl dark red} as a function of time. The four panels ({\sl from top to bottom and left to right})
corresponds to different stages of the winding of two walls around a north-east oriented filament. Once the dark matter has joined the filament, it heads towards the bottom left part 
of the panel in the direction of a more massive node of the cosmic network. 
This process is best seen dynamically on {\em\tt
http://www.iap.fr/users/pichon/spin/}
}
\label{fig:visual-tracerDM}
\end{figure*}

The measurements of section~\ref{sec:spin-fil}  strongly suggest that the spin direction of dark matter halos is connected to the cosmic web. Indeed,
these dark halos are embedded in large-scale cosmic flows induced by the successive formation of walls, filaments and clusters: matter escapes from the voids to the walls then to the filaments before flowing along this latter direction towards the nodes \citep{zeldovich70,bkp96,pichonetal11}.  In this framework, let us now argue  that
the first generation of halos are formed  during the {\sl winding of  walls around filaments}, which  provides them with a spin parallel to this direction (and whose amplitude is proportional to the relative impact parameter of the  two  walls). 
   Conversely, the later generations of halos form by mergers along the filaments i.e.  in the direction parallel to the mean flow \citep[as was first pointed out by][]{b6} and therefore acquire a spin perpendicular to the filaments (if the orbital angular momentum which is converted into spin during the merger dominates). 
In this scenario, the transition in mass measured in fig.~\ref{fig:mass} in fact reflects a trend in merging generation\footnote{Technically it was not possible to build merging trees a posteriori on the  Horizon 4$\pi$ simulation as it  would have required too many snapshots and thus too much disc space; so we did carry out the tests described in 
Section~\ref{sec:treemaker} on a smaller simulation
to confirm the relevance of merger generation as the key parameter.}. Note that this behaviour occurs on several scales simultaneously; this multi-scale signal is probed by varying the smoothing scale used to define the filaments in Appendix~\ref{sec:smoothing}.

As a first check for this hypothesis, the typical distance of DM halos from filaments is computed as a function of  their mass. The more massive halos are  typically found closer to the filaments (on average at 0.7 Mpc/h) than low-mass halos (found on average at 2 Mpc/h from the core of the filaments), which is consistent with our  assumption because it implies that high-mass halos have reached the center of the filaments, where they are more likely to merge in the direction of the flow. Conversely, low-mass halos (for which large-scale dynamics has been frozen in at an early stage) are found further from the core of the filaments where they are less likely to merge. This is qualitatively consistent with fig.~\ref{fig:peaks}, which shows the distribution of massive halos within a slice of the simulation.

\subsection{Winding up of dark matter flows}
\label{sec:DM}

Let us have a look at fig.~\ref{fig:visual-tracerDM}, which displays the temporal evolution of dark matter particle trajectories in the vicinity of a dark matter filament.
This simulation has the special feature that the small-scale modes were erased from the initial conditions in order to facilitate visualization of the large-scale flow, see \cite{pichonetal11} for details.
We refer to this simulation as a ``smoothed'' simulation. Here most of the flow is in fact embedded in a large wall parallel to the plane of the figure. 
The dots of different colours (from {\sl yellow} to {\sl blue} via {\sl dark red}) correspond to the position of the same dark matter particle at successive time steps, and thus allow us to visually follow 
the trajectory of DM particles. The top left panel corresponds to a snapshot somewhat before the flow has significantly shell-crossed around the north-east filament. 
The DM particles are plunging towards their filament, while flowing within the two walls. 
On the top right panel, some level of shell crossing has occurred
in the north east part of the filament, and the corresponding particles have started inflecting their trajectories to wind up around the locus of that filament. 
Since these particles typically have a non zero impact parameter relative to 
the center of mass,
 as they wind up, they convert their orbital motion into spin while generating a virtualized structure.
Later on, (bottom left panel) this structure sinks along the north-east filament towards a more  massive clump (off field).
Meanwhile, the process of DM winding  from the walls around the main filament continues, and feeds (as a yellow trail)
 the dark matter halo along its  current spin axis (which is aligned with 
the axis of the filament). Finally on the bottom right panel, another such halo has formed further down the filaments, and we can anticipate that their upcoming merger will
lead to a structure whose spin's direction will be a mixture of their initial spin, and the spin perpendicular to their relative orbital plane.
In section~\ref{sec:hydrodynamics} below we will revisit this scenario using hydrodynamical simulations, which allow us to visually identify the spin of forming galaxies.

\subsection{The progenitors of dark halos via merging trees}
\label{sec:treemaker}
In order to understand the  previously described mass transition and its redshift evolution, we used the code {\tt TreeMaker} \citep{b37}
to track down the progenitors (both dark halos and unresolved flow) of given halos in conjunction with their spin orientation relative to the nearest filament.
TreeMaker involves two steps: first halos are identified using a halo finder -- in our case a FOF algorithm \citep{b27, zeldo82, davis85}, while the
properties of these structures (mass, angular momentum ...) are measured. As a second step, the individual DM particles which belong to each
halo are tracked back in time so as to build a merging history tree  which regroups all of its progenitors and their properties as a function
of time.
A relatively large number of merging trees  were computed and the direction of the spin of the progenitors relative to their host filament was calculated.

After visual inspection of  a  subset of those merging trees, it was found that: 
i) the high-mass halos (i.e above the critical mass) with a spin perpendicular to their filament tend to have a similar history:
 they often acquire a significant amount of mass via a major merger, which is accompanied by a significant
 spin adjustment from a direction initially aligned with to a direction mostly perpendicular to their filament (and aligned with $\mathbf{e}_2$, see Appendix \ref{sec:shear});
ii)  in contrast, low-mass halos (i.e below the critical mass) are not the result of major mergers; often no mergers at all are found at the mass resolution of the N-body simulations; those who have a spin parallel to the filament seem to have acquired this  spin direction at a time (the so-called formation time) when they have acquired most of their mass by diffuse accretion.

\begin{figure*}
\includegraphics[width=0.9\columnwidth]{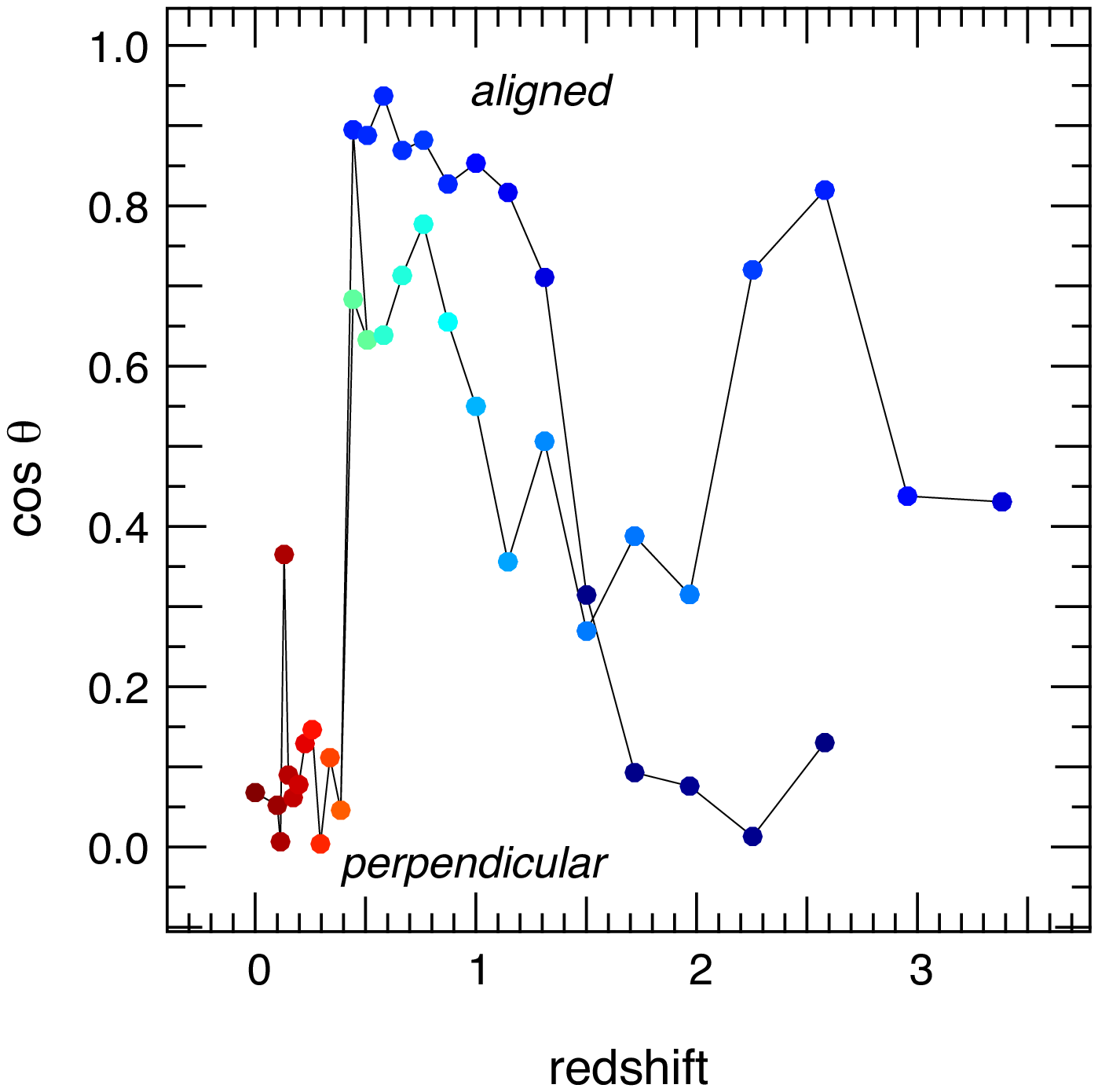}
\includegraphics[width=1.025\columnwidth]{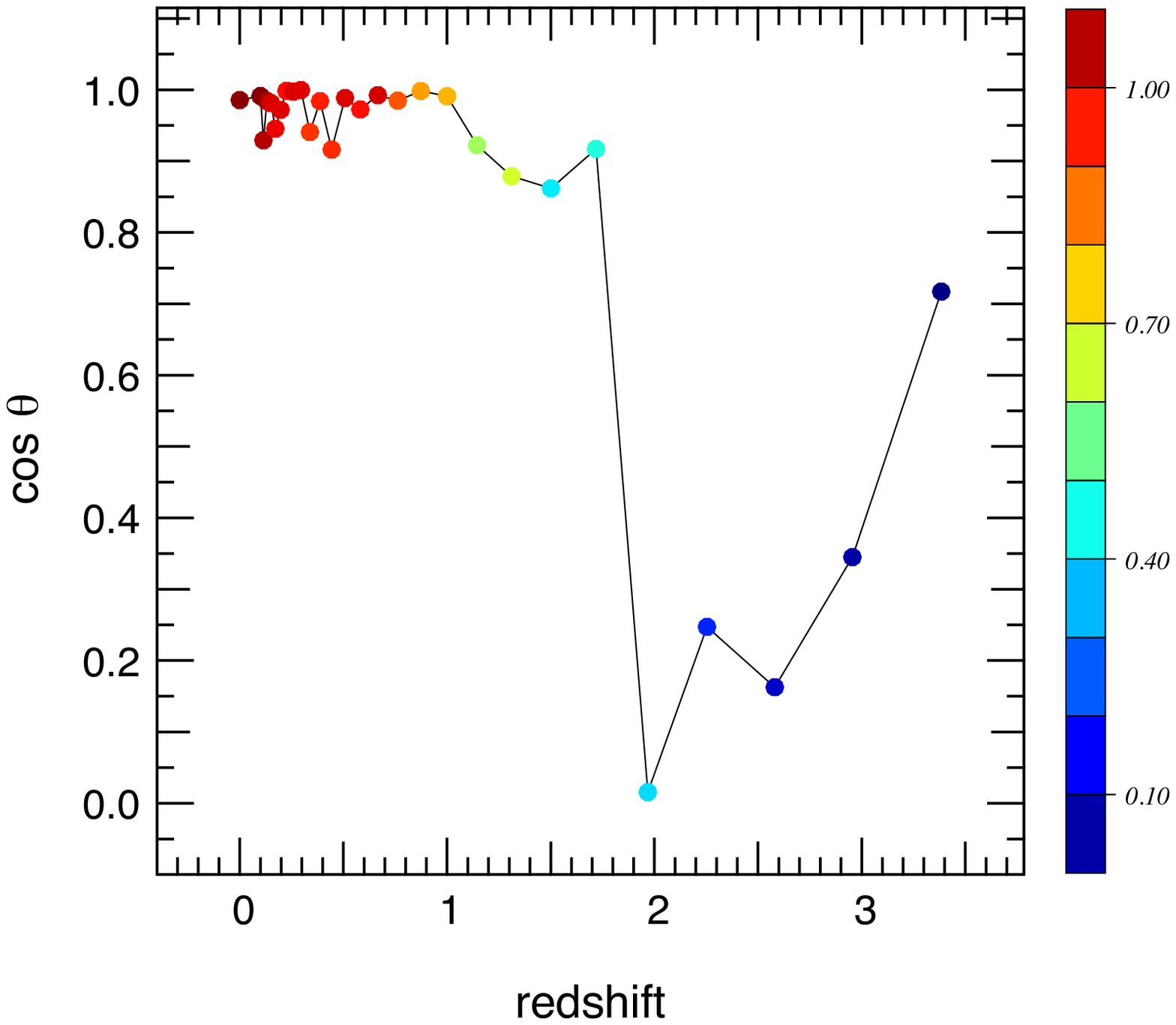}
\caption{{\sl Left}: merging tree of a high-mass halo ($8\cdot 10^{12}M_{\odot}$ at redshift 0). The various colors correspond to different mass fractions (relative to $z=0$). The vertical axis correspond to the angle between the host filament and the spin of this halo and horizontal axis to the redshift. Here, the spin becomes perpendicular to its filament after an important merger at redshift $\sim$0.5.
{\sl Right}: merging tree of a low-mass halo ($2\cdot 10^{11}M_{\odot}$ at redshift 0).  Here, the spin becomes suddenly parallel to its host filament when the halo acquires most of its mass by accretion between redshift one and two.  } \label{fig:MT1} \label{fig:MT2}
\end{figure*}

This behaviour is quite generic as we observed it for a few tens of randomly chosen halos.
It is illustrated in fig.~\ref{fig:MT1} where merging trees of two different halo (extracted from one of  the  $50$ Mpc/h $\Lambda$CDM simulations) are shown: the evolution of the angle between the closest filament and the spin of a given halo is plotted as a function of the redshift; the colors encode the fraction of mass of the progenitor with respect to the final halo mass. The right panel corresponds to a low-mass halo ($2\cdot 10^{11}M_{\odot}$ at redshift 0 corresponding to more than 300 particles) which forms at redshift $z \simeq 1.5$ (when it has already acquired more than one half of its mass) and suddenly acquires a spin parallel to its closest filament at a high redshift ($\simeq 2$) close to its formation time. This halo does not undergo any significant merger afterwards.  Conversely, the left panel provides the merging tree of a high-mass halo ($8\cdot 10^{12}M_{\odot}$ at redshift 0)  which forms at lower redshift ($\simeq 0.4$) as the result of a major merger between two less massive halos. This event corresponds exactly to the time when it acquires a spin perpendicular to its closest filament. What is striking here is the clear flip of the spin direction: the two progenitors have a spin aligned with the filament (this spin is acquired at higher redshift $\simeq 1.5$) and their merger makes the spin of the resulting halo become perpendicular to it. These two examples of trees  are characteristic of  how halos below and above the critical mass $M^s_{\rm crit}$ form and acquire spin. Note that we actually observe a large dispersion of the histories around this mean behaviour. 

\begin{figure}
 \includegraphics[width=0.95\columnwidth]{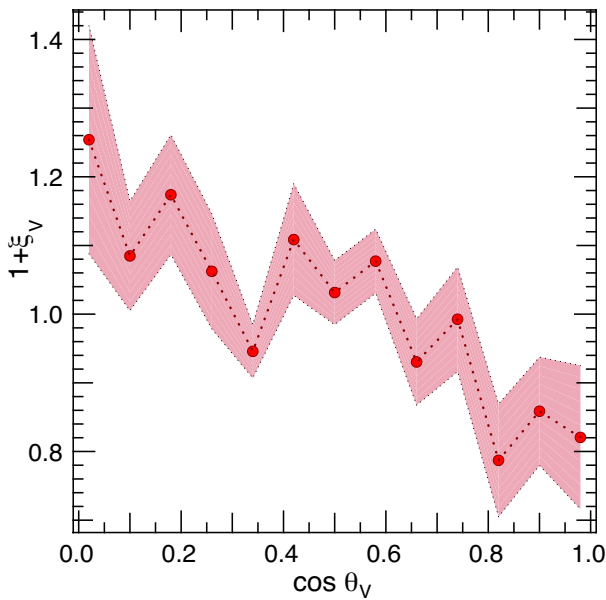}
 \caption{Excess probability of   the velocity-filament alignment just before formation time for a sample of 15 low mass halos ($M<M_{\rm crit}$).
 The average orientation over these 15 halos is plotted with a dotted line, while the 1-$\sigma$ deviation away from that mean is displayed in pink.
 The progenitors' velocities before formation lies preferentially in a plane perpendicular to the closest filament (see also figs.~\ref{fig:MT2} and~\ref{fig:walls}). 
 }
 \label{fig:velocity}
\end{figure}

To check the statistical robustness of this scenario for low-mass objects,  let us identify a preferred plane of motion at  formation time.The following simple test was implemented: {a set} of low-mass halos ending with a spin parallel to their closest filament are randomly chosen and only halos for which a time of significant accretion (i.e. their formation time) can be determined are retained (this represents at least one third of our sample). 
Their particles are  traced back one time step before their formation time to quantify the relative orientation of their velocities
 compared to the filament's direction. The  excess probability distribution of alignment, $1+\xi_V$ is displayed in fig. \ref{fig:velocity} and shows that their velocities (before formation) are more likely to lie  perpendicular to the filament (in particular it is found that $\left\langle \cos \theta_V\right\rangle\simeq 0.47$), which is in  agreement with the  scenario (see also fig.~\ref{fig:walls})).

 In order to assess the statistical relevance of the scenario for massive halos (i.e above the critical mass at redshift zero), merging trees can also used to determine their most recent merger.
 The mean cosine of the angle between these halos and their host filament just after merging is computed and compared to the mean cosine of the angle between the closest filaments and the spin of the progenitors just before merging. This test yields $\left\langle \cos \theta\right\rangle\simeq 0.51$ before and $0.47$ after the last merger, which is fully consistent in amplitude with the mean cosines found for the Horizon $4\pi$ simulation (for which for instance $\left\langle \cos \theta\right\rangle\simeq0.510$ (resp. $0.479$) for $M\simeq 10^{12}M_{\odot}$ (resp. $M\simeq 10^{13}M_{\odot}$)). The 1-$\sigma$ error  on the mean is about $\pm 0.02$, given the size of this sample (only $\simeq 200$ halos massive enough in this simulation). This statistical test is in agreement with the fact that massive halos acquire a spin perpendicular to their host filament because of mergers.

 Altogether, we are now able to reconstruct the history of spin acquisition by DM halos following the large-scale dynamics.
The less massive objects are born by accretion at high redshift during the winding of the walls into filaments. This process generates a spin aligned with the filaments. 
Most of the halos of low mass at $z=0$ are now formed, and their spin will not change much because they have already acquired most of their mass while future  accretion will not be important enough to have a strong impact of their spin direction. This behaviour is illustrated on fig. \ref{fig:MT1} (right panel) where a halo forms and acquires a spin parallel to the direction of the forming filament, which then remains in the same direction during accretion until redshift zero.

At lower redshift, filaments are collapsing and thus create a flow along their direction in which the more massive halos form by major mergers. During this process,  these massive halos acquire a spin which is the superposition of the spin of their progenitors and the orbital spin from the merger. As the motion is along the filament, this orbital spin is in the plane perpendicular to it. This process was shown on fig. \ref{fig:MT2} (left panel):  the spin of the (less massive) progenitors are parallel to the filaments and their merger within the filaments induces a more massive halo whose spin is now perpendicular to the host filament. For these more massive dark matter halos, a competition between orbital spin and intrinsic spin during the merger process has just been highlighted and explains the resulting orientation. Indeed, their spin are not randomly distributed in the plane perpendicular to the filaments but  are shown to be correlated with one particular eigen-direction of the large-scale tidal tensor. 
This issue is fully addressed in Appendix \ref{sec:shear}. 

 It is interesting to compare the picture described above with \cite{peirani04} who focused on the spin magnitude (instead of its direction). 
 These authors claimed that  spin acquisition was dominated by merger events rather smooth accretion; 
  fig. \ref{fig:MT1} suggests that this holds true for the spin's direction of halos more massive than the critical mass $M^s_{\rm crit}$ as  well.

\subsection{Visual inspection using hydrodynamics}
\label{sec:hydrodynamics}

 \begin{figure*}
   \includegraphics[width=\columnwidth]{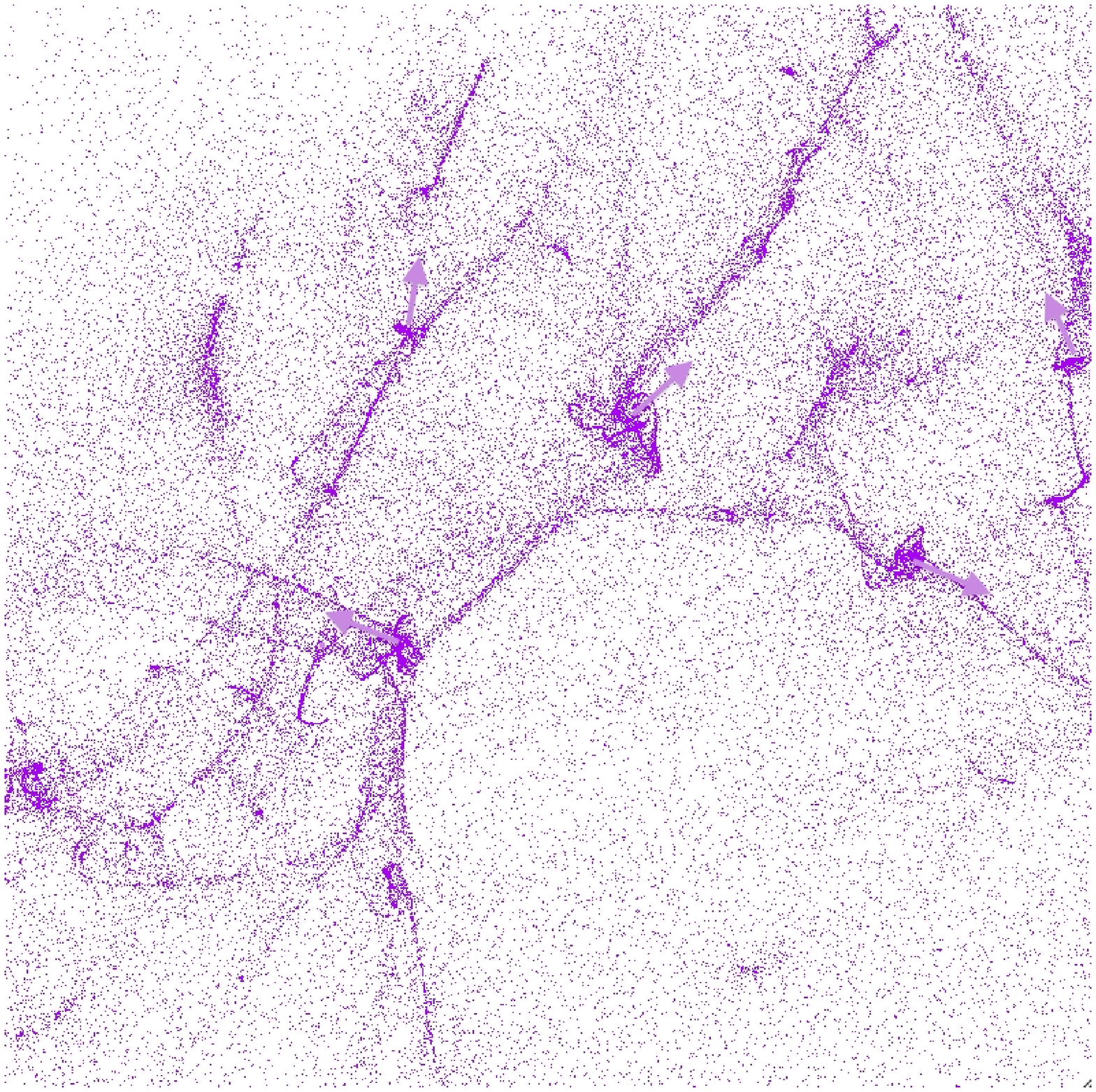}  	
    \includegraphics[width=\columnwidth]{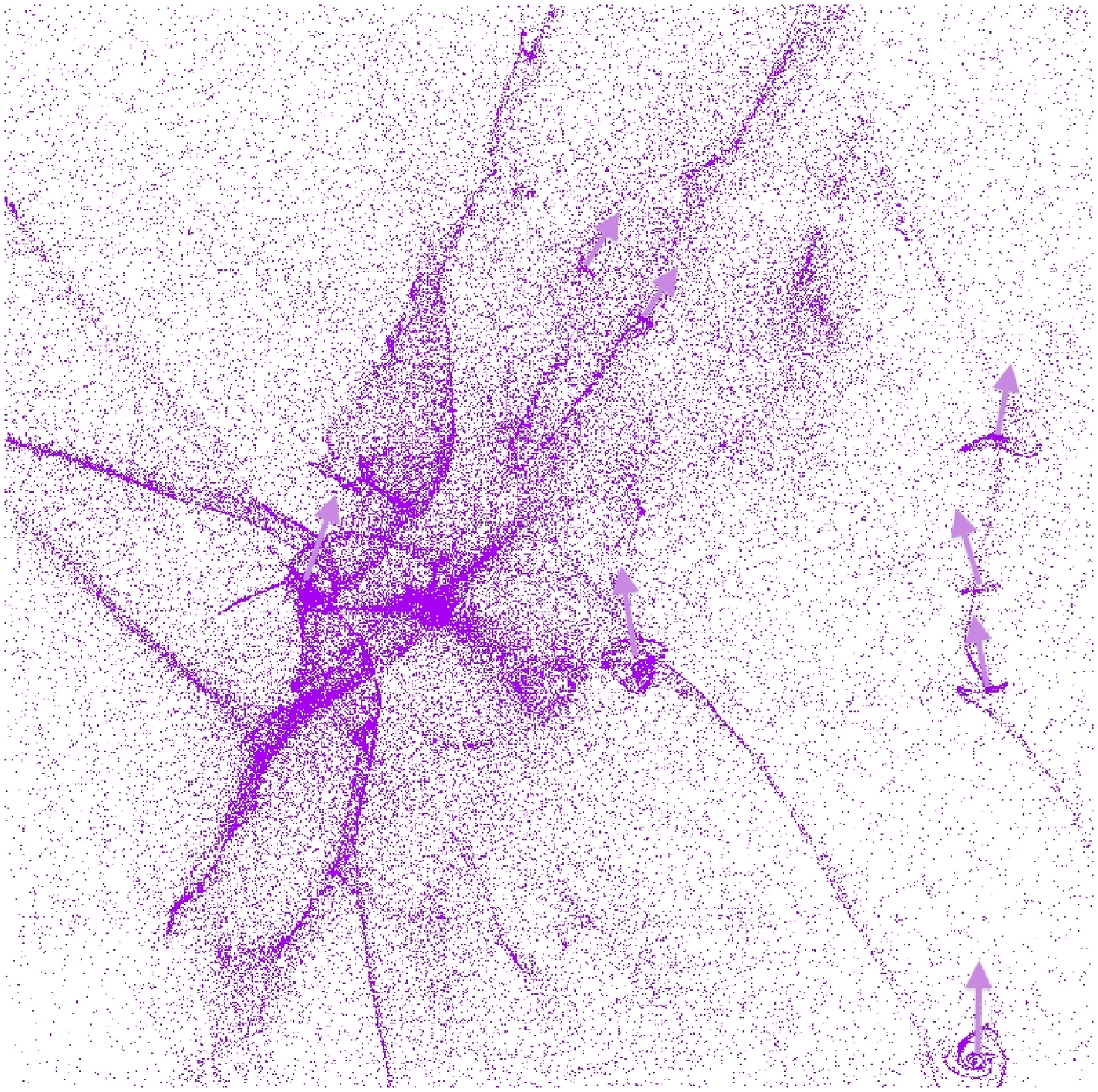}  	
      \includegraphics[width=\columnwidth]{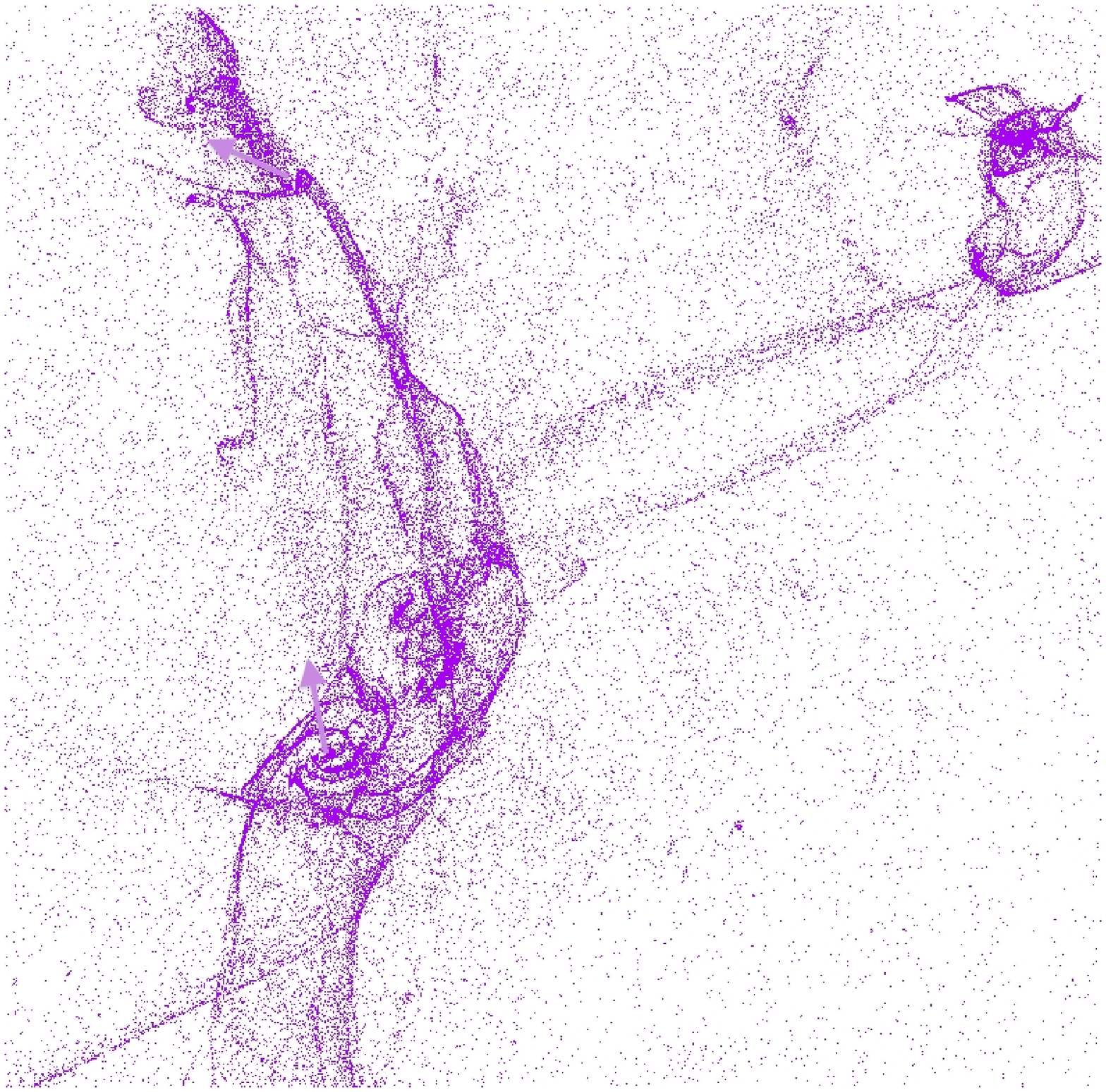}  		
     \includegraphics[width=\columnwidth]{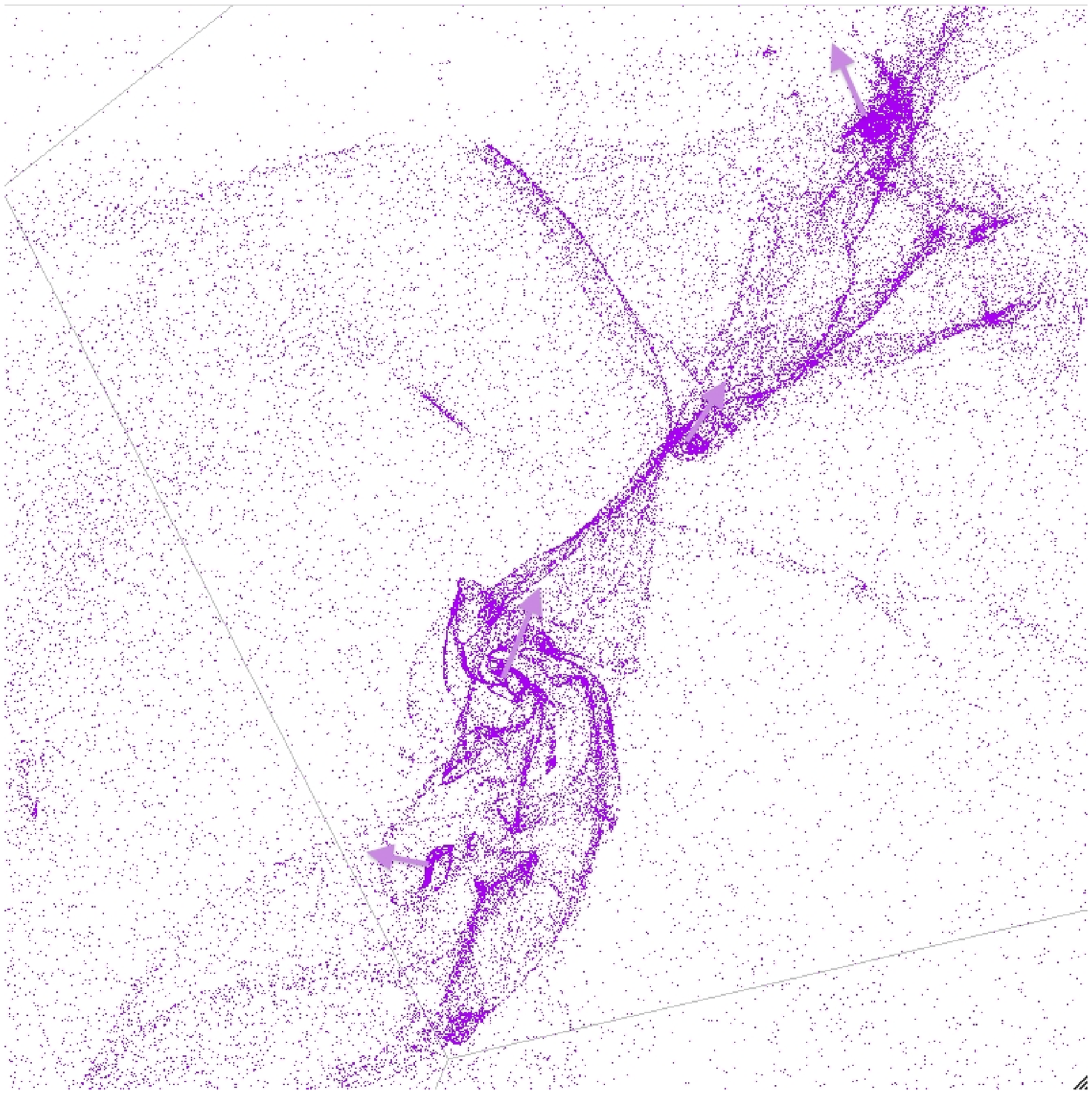}  		
 \caption{Distribution of hydrodynamical tracer particles (at $z\simeq9$, top panels, and $z\simeq8$ bottom panels). The web-like filamentary structure ({\sl top left and right panels}) of the gas distribution which ends up in the bulge of a unique galaxy at later times 
      is quite intricate, though one main wall in which the largest filament is embedded dominates (seen more clearly edge on, in the top right panel). 
      Note the disc-like features with a spin parallel to the filament (represented qualitatively as an arrow perpendicular to the disc). 
      {\sl Bottom panels:}  zoom in at later stage, to visualize the process of a merger along the filament, 
      before ({\sl bottom left panel}) and during  ({\sl bottom right panel}) the merger.  The spin of the merger remnant is a combination of the orbital 
      angular momentum and the initial spin of the progenitors; it can therefore depart from the direction of the host filament. Note  also the ribbon structure of filaments which corresponds to the locus of the second shock. 
      Visual inspection suggests these ribbons become broader with time (as predicted by  \protect\cite{pichonetal11} as they advect  larger and larger amount of 
angular momenta from the outskirt of the gravitational patch) and tend to lie perpendicular to the main wall; as they reach the protogalaxy, they twist rapidly  on outer shells and build  up its outskirt.
}
\label{fig:visual-tracer}
\end{figure*}
%
\begin{figure*}
    \includegraphics[width=\columnwidth]{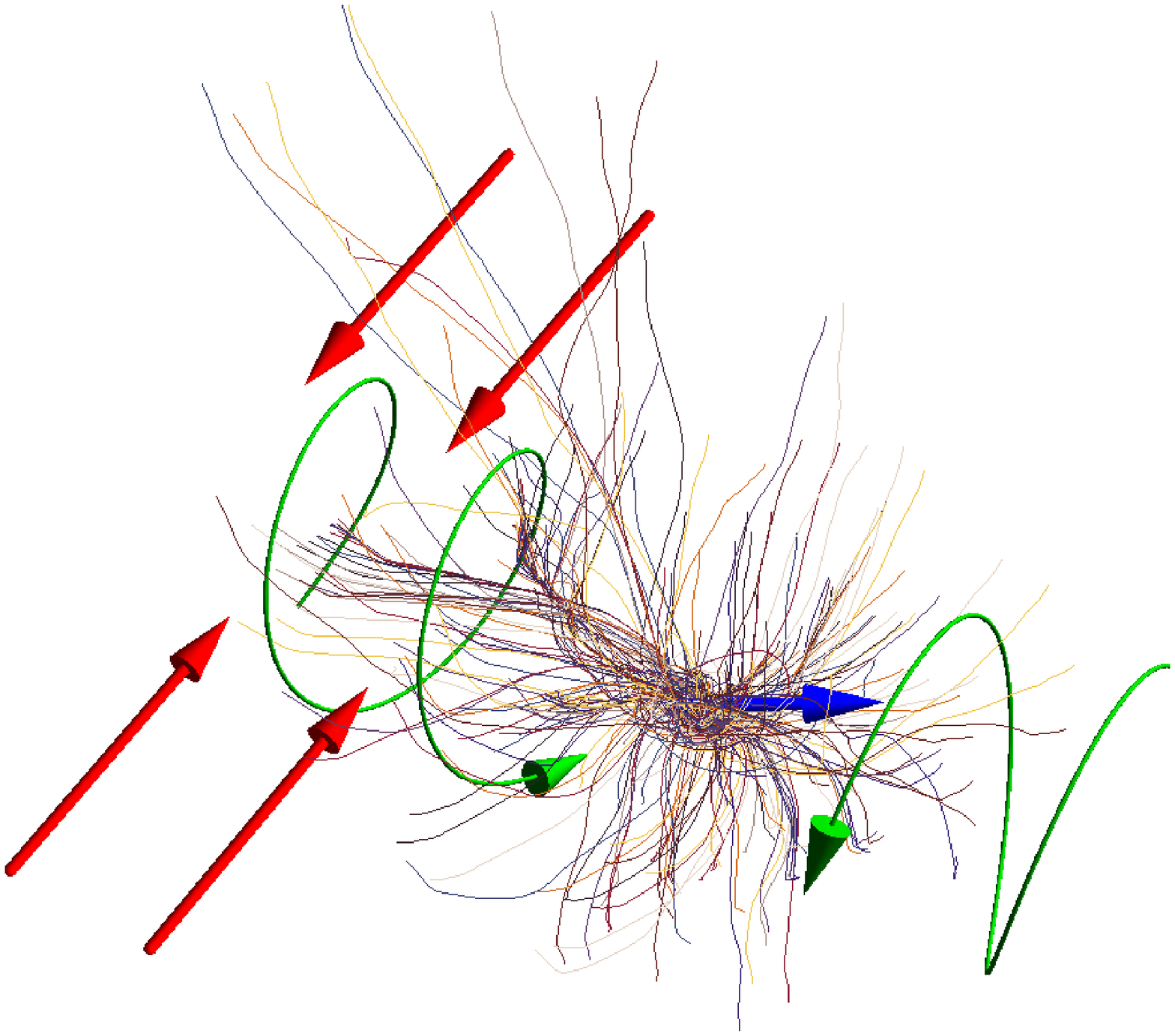}  	
      \includegraphics[width=0.8\columnwidth]{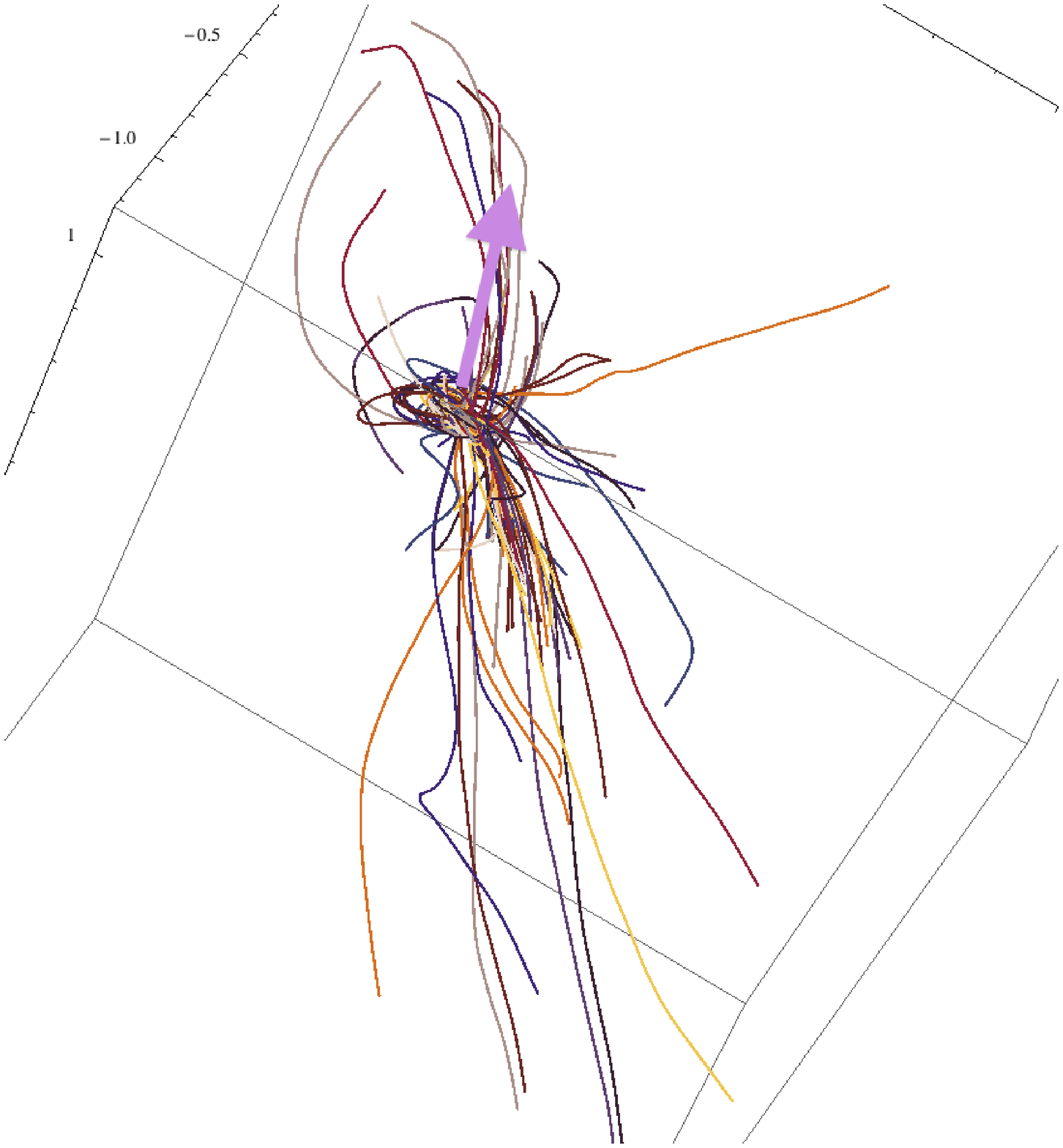}  		
  \caption{Trajectories of tracer particles in the outer region ({\sl left panel})  and 
     inner region ({\sl right panel}). On large scales, the gas departs from voids, flows along the walls ({\sl red arrows}) and wind up in filaments along well defined ribbons ({\sl green arrows}), forming low-mass halos. In the inner region of the tracks,  the flow is indeed along the filament and spiraling in into a disc whose axis is roughly aligned with the filament.}
\label{fig:visual-tracer2}
\end{figure*}
For illustrative purposes, let us now turn to a high redshift hydrodynamical simulation, as the cold gas that we will analyze here follows more closely the caustics of the 
cosmic web than the dark matter, and thus provides a clearer visual impression of the process of  wall winding and spin acquisition
 (in contrast to collisionless DM, cold gas does not undergo shell crossing but shocks and loses degrees of freedom -- its  motion perpendicular to the shock). 
We are not concerned here with how much respective angular momentum gas has with respect to dark matter acquired, but only use hydrodynamics as a proxy for pinpointing
more accurately the loci of shell crossing and identifying the spin axis of galaxies. As we will argue later, it is also of interest to consider in parallel the 
environment of  low-mass high-redshift  and high-mass low-redshift halos.
 Following  \cite{Dubois2011} we  use tracer particles of the gas in a cosmological hydrodynamical simulation (which is described in Section \ref{sec:ap-hydro})
  to illustrate this winding of walls and the loci and orientations of galaxies.

 Fig.~\ref{fig:visual-tracer} represents 
the web-like filamentary structure of the gas at $z=9$ in a field of size  approximately 50 kpc$/h$ across;  the  ensemble of tracer particles initially makes up a
sheet-like structure   with a dominant filament embedded in it. Note that these  tracer particles represent a biased subset of all tracers as they are chosen so as to 
end up within the bulge of the main galaxy of this zoom simulation at some later stage. The tracer particles flow  from this sheet into the filaments where they form
"protogalaxies". The  gas  thereby  typically has a non zero impact parameter relative to the filament  and protogalaxies thus acquire a spin parallel to the filament in which they form
(see arrows on the figure). 
These young galaxies then migrate along their filaments and merge with other galaxies.
The spin of the merger remnant is  a combination of the orbital angular momentum of the collision and the initial spin of the progenitors; it can therefore depart from the direction of 
the host filament. We provide an animation which allows to better see this at {\em\tt
http://www.iap.fr/users/pichon/spin/}.  Fig.~\ref{fig:spin-yohan} illustrates quantitatively the visual impression of 
fig.~\ref{fig:visual-tracer} while measuring the spin of the circum-galactic discs (between $0.1 R_{vir}$ and $0.25 R_ {vir}$).

In fig.~\ref{fig:visual-tracer2},  a small subset of these tracer particles are randomly chosen and followed for a while from early (left panel) to late times (right panel).
On large scales (at early times), we note that the flow is indeed dominated 
by the winding up of matter from the main wall around the main filament. The tracers' trajectories start perpendicular to the filament within the walls.
As they reach the filament, they take a sharp turn, losing their transverse motion and flowing along the filament (left panel).
 In doing so, since the laminar flows on opposite sides of the wall will typically have different impact parameters, they generate a spinning structure whose axis 
 will be aligned with the filament.
 This  spaghetti structure converges into a quite narrow and elongated plait on either side of the forming disc.
Given its orientation the induced disc will therefore advect secondary infall at its periphery preferentially along its spin axis (as both the galaxy and its upcoming secondary infall were assembled via the same winding process). 

In a nutshell, the cold gas dynamics  of large-scale cosmic flows  provides  a much clearer illustration of the scenario we outlined for the formation of halos
with a spin mostly aligned with the local filament. We do not discuss any further  the properties of the alignment of the  disc w.r.t. the filamentary structure given 
that there exists many caveats (though see Section~\ref{sec:speculations} below and Appendix ~\ref{sec:ap-hydro}).

\section{Conclusions \& prospects}
\label{sec:conclusion}
In this paper, the Horizon 4$\pi$ N-body simulation was used to investigate the correlations between the spin of dark matter halos and their large-scale environment. 
For filaments defined over a smoothing scale of $5 h^{-1}$ Mpc, a statistically significant signal was detected, indicating that the orientation of the spin of dark matter halos is sensitive to the cosmic environment.
A mass dependence of this signal was also robustly established: low-mass halos are more likely to be aligned with large-scale filaments (with an excess probability of 15~\%) 
whereas more massive halos tend to be perpendicular to these (with an excess probability of 12 \%). 
  The mass transition was found to be redshift dependent and to vary  like $M_{\rm crit}(z)\approx M_0 \cdot (1+z)^{-\gamma}$ with $M^s_0 \simeq 5(\pm 1) \cdot 10^{12} M_\odot$ and
  $\gamma_s= 2.5 \pm 0.2 $.
  This critical mass is also found to increase with the smoothing length (Appendix~\ref{sec:smoothing}). These results are in agreement with those presented in Appendix \ref{sec:shear}, which are derived using
  the more classic approach of considering excess alignment of the halo spin with the tidal tensor eigen-directions. Since the tidal tensor probes larger scales than the skeleton of the 
  density, the mass transition is detected for a larger mass $M^t_0 \simeq 8(\pm 2) \cdot 10^{12} M_\odot$, which scales slightly differently with redshift, i.e.
  $\gamma_t = 3 \pm 0.3$ .
Both redshift evolutions are roughly consistent with that of the formal non-linear mass scale.

\subsection{Discussion }
\label{sec:discus}

 This unambiguous result confirms and quantifies  recent findings \citep{b24,b10,b14,b16}.   It is also consistent with \cite{b7} who found that the spin of dark haloes tend to be perpendicular to filaments. Indeed,
    weighting our statistics by the spin magnitude (which corresponds to their strategy), we recover  their results. \cite{b18} found
  the same result, but  did not have enough statistics to probe the signal above the critical mass. The study of \cite{b23} focused on halos above $10^{14} h^{-1}M_{\odot}$, i.e. above the critical mass at redshift zero, which is why, in agreement with this work, they found a trend for these halos to have a spin perpendicular to their closest filament. In contrast, \cite {b33} considered halos with mass around $10^{11-12}M_{\odot}$,
     i.e below the critical mass and thus found an alignment of the spins with the filaments, again in agreement with this work. 
This paper also confirms the very recent findings of \cite{libeskind12} who claimed that low-mass haloes (around $10^{10-11}M_{\odot}$) tend to be aligned with the filaments and to lie in the plane of the walls.

Note that  Appendix~\ref{sec:shear} predicts a strong trend for the spin vector
of DM halos to lie in the plane of large-scale walls. 
Such a signal was claimed to have been detected in observations  by \cite{b13,trujillo06,navarro04} and should be re-investigated in view of this paper's predictions.  
Furthermore, observers should now also be able  to investigate the spin-filament correlations in a way directly comparable to the theoretical predictions for DM presented in this paper (fig. \ref{fig:mass}). Indeed, the code DisPerSE \citep{sousbie11} which identifies filaments using persistence, can accurately deal with discrete and sparse datasets and should 
provide a good estimator for the direction of  {\sl observed } filaments.

The time evolution of the angular momentum of individual dark matter halos obtained by following their progenitors using merger trees  suggests that the spin 
direction results on the one hand from the winding of the walls into filaments (first generation, low-mass halos) and on the other hand from significant mergers occurring along those 
filaments (second generation, more massive halos). More specifically, the arguments we developed throughout the paper strongly suggest that
the measured correlations can be understood as a consequence of the dynamics of large-scale cosmic flows. 
Indeed, low-mass halos mostly form at high redshift within the filaments generated by colliding/collapsing walls. Such a process naturally produces a net halo spin parallel to the filaments.
In contrast, high-mass halos mainly form by merging with other halos along the filaments at a later time when the filaments are themselves colliding/collapsing. Therefore
they acquire a spin which is preferentially perpendicular to these filaments. Visual examination of 'smoothed' dark matter and hydrodynamical simulations lend extra support to this picture (see also Appendix~\ref{sec:ap-hydro}). 
 Measurements of  the orientation of the spin relative to the eigenvectors of the tidal tensor are  also consistent with such a scenario provided one takes into account 
 the fact that they probe typically 
 larger scales of the density field. 
 
 From the point of view of a large-scale filament, 
  most low-mass halos are formed early from patches
  that are part of  a planar, flattened inflow of  matter
  onto that  filament. For gaussian random fields, 
  the tidal tensor  in such patches is correlated with the filament's direction \citep[via the shape parameter $\gamma$,][]{Pogosyan2009},
  resulting in the preferential alignment of the spin of such halos
  along that filament.
  This process is related to the theoretical predictions of \cite{b22} who demonstrated, using the Zel'dovich approximation, how 
    vorticity was generated during the first shell-crossing. This vorticity will  lie in the plane of the forming  walls.
Extending their predictions while focussing now on a 2D flow, we speculate that
secondary shell-crossing will lead  to the formation of   vortices   aligned with the forming filament  \cite[see Figures 5 and 7 of ][a possible  section perpendicular to the axis of the filament]{b22}\footnote{Note however that such a typical caustic should have inherited some level 
of asymmetry \citep{pichonetal11}, which could imply that one vortex dominates.}. In turn these vortices could account for the spin of  protogalaxies, 
as was suggested by the referee of that paper.

The excess probability in figs.~\ref{fig:mass} and  \ref{fig:shear} lies at the 15-40~\% level. As such it mainly reflects a residual trend of coherence inherited from the large-scale
cosmic environment in which halos form. 
This does not preclude the multi-scale hierarchical clustering process to erase part of this more orderly dynamics.
For instance, clustering and merging on smaller scales will in part perturb the large-scale ordered ribbons (as is already visible on the bottom panels of fig.~\ref{fig:visual-tracer}).

{ The trend in fig.~\ref{fig:mass} is found  on the
one hand in these measurements at the high end of the mass function at redshift zero (whose dynamics is only mildly non-linear), and on the other hand for the gas (a proxy for DM caustics) at very high z for lower mass  galaxies (again not very far from linear dynamics at this epoch).  Hence we found in both r\'egime that some imprint of the large-scale structure geometry and dynamics is directly
responsible for the spin of the forming object and its post merger transition.
 }
In this paper we tried and explained the  origin of the statistical signal, but we do not argue that large-scale dynamics dominates the non-linear regime of galaxy evolution on an individual object basis. 

A remaining task involves understanding  in details the redshift-dependence of  the transition masses $M_{\rm crit}$ described by equations~(\ref{eq:trendz}) and~(\ref{eq:trendz2}) (and also its dependence with the smoothing length, which might allow us to identify a scale at which the (anti-)alignment is strongest).  
It would also clearly be of interest to further quantify  (using larger samples) 
the findings of \cite{hahn10} regarding the alignment of the stellar disc/circum-galactic medium with respect to the LSS
and investigate how these results depend on sub-grid physics and feedback processes.

Let us conclude this paper by some speculations about what these results imply specifically for the process of galaxy formation at high redshift.

\subsection{Implications for galaxy formation}
\label{sec:speculations}
 
We are now in a position to {\it speculate} about the implication of our findings for high redshift galaxy formation.
In the view of the robust measurements of Sections~\ref{sec:fil-spin} and Appendix~\ref{sec:shear} and the visual inspections of Section~\ref{sec:discussion}, it appears  that i) galaxies form preferentially along filaments and ii) that their internal dynamics (hence their morphology)  inherit important features from this anisotropic environment.
The first point is backed by the distribution of filaments at the virial radius \citep[Appendix A of ][and indirectly by fig.~\ref{fig:peaks}]{pichonetal11,danovichetal11},  the second point by the measurements 
reported in this paper.

Indeed, filaments can be thought of as the loci of constructive interferences from the long-wavelength modes of the initial power-spectrum.
On top of these modes, constructive interferences of high frequency modes produce peaks which thus get a boost in density that allows them to pass the 
critical threshold necessary to decouple from the overall expansion of the Universe, as envisioned in the spherical collapse model \citep{Gunn1972}. 
This well known biased clustering effect has been invoked to justify the clustering of galaxies around the nodes of the cosmic web \citep{WhiteTullyDavis88}.
It also  explains why galaxies form in filaments. In walls alone, the actual density boost is not sufficiently large to trigger galaxy formation.
We therefore argue here that, statistically, the main nodes of the cosmic web are where galaxies migrate,
not where they form;  galaxies generally form  while reaching  filaments from walls. They thus inherit the anisotropy of their birth place
as spin orientation. During migration, they may collide with other galaxies/halos and erase part of their birth heritage when converting orbital momentum into spin via merger.

\begin{figure}
\center    \includegraphics[width=0.95\columnwidth]{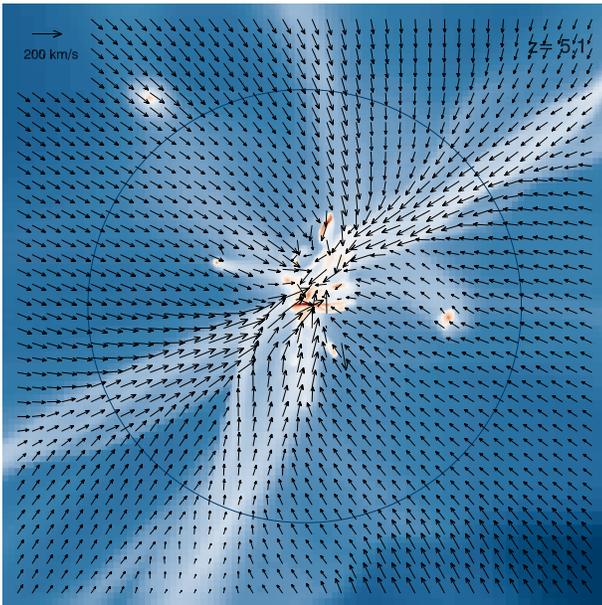}  	
     \caption{a meridional projection through the velocity flow and gas density around the {\nut} galaxy at  redshift 5.1 \protect\citep{kimmetal11}.
     The disc plane is along the horizontal axis, the circle marks the virial radius (17.6 kpc).
      Note the ribbon-like
      cold-flows 
     seen directly in the gas density and the velocity flow, which hit the galactic disc roughly along its spin axis.
     Note also how the gas takes a sharp turn when it reaches the ribbons.}
\label{fig:visual-nut}
\end{figure}

Recently,  \cite{pichonetal11} showed how the cold gas
drains out of  the prominent voids in the cosmic web, into 
sheets  and filaments before it finally gets accreted onto 
dark matter halos. 
Interestingly, the imprint of  the larger-scale pancake-structure of the typical cosmic web around a filament and a peak allows us to be more specific about the geometry of this process.
Indeed, one of the striking  features of figs.~\ref{fig:visual-tracer} and.~\ref{fig:visual-tracer2} (probably best seen in the animation online) are these ribbon-like caustics which feed the central galaxy along its spin axis 
from both poles.  Generically the gas inflow in the frame of the galaxy is double helix-like along its spin axis; 
this is mostly wiped out in DM (and hardly visible in fig.~\ref{eq:trendz}) because of shell crossing,
 but quite visible for the gas.  These ribbons are generated via the same winding/folding process as the protogalaxy, and represent the dominant source of secondary filamentary infall
 described in  \cite{pichonetal11}, which feeds the newly formed galaxy with gas of well aligned angular momentum (whose direction was set by the impact parameter offset of the 
 two neighbouring  walls, which can in turn be attributed to the dissymmetry of the four neighbouring  voids).  As such, the larger-scale geometry of the LSS 
 (which biases the formation process) squashes the  average neighbourhood of a peak (6 saddles, 8 voids, 12 walls), into a simpler effective geometry (one wall and one embedded filament dominating). Formally,
the most likely ``crystal" of the universe -- subject to the constraint of collapse along two axis on larger scales, differs from the azimuthally-averaged cubic centered crystal found in Pichon et al. ({\sl in prep.}): it is quite flattened and dominated by one ridge \citep{pichonetal11,danovichetal11,Dubois2011}.
Note that the gas flowing roughly parallel to the spin axis of the disc along both directions will typically impact the disc's circum-galactic medium and shock once more (as it did when it first reached the wall, and then the filaments, forming those above mentioned ribbons), radiating away its vertical momentum (see fig.~\ref{fig:visual-nut} and {Tillson et al, in prep.}). 

Our speculations here have focussed on a two-scale process. Given the characteristics  of $\Lambda$CDM hierarchical clustering 
one can anticipate that this process occurs on several nested scales at various epochs - and arguably  on various scales at the same epoch\footnote{The scenario we propose for the origin of this signal is, like the signal itself, relative to the linear scale involved in defining the filaments and as such, multi-scale. It will hold as long as filaments are well defined in order to drive the local cosmic flow.
}. In other words,
one expects smaller-scale filaments are themselves embedded in larger-scale walls (as discussed in Appendix~\ref{sec:shear} to reconcile our excess alignment with 
the eigenvectors of the tidal tensor). The induced multi-scale anisotropic flow transpires in the scaling of the transition mass with smoothing, as discussed in Appendix~\ref{sec:smoothing}.

Another issue would be to estimate for how long  this entanglement between the large-scale dynamics and the kinematic properties of high redshift  pervades.
Indeed, \cite{ocvirketal08} have shown that at lower redshift, the so-called hot mode of accretion will kick in; 
how will  hot flows  wash out/disintegrate  these ribbons?  Given that they locally reflect the large-scale geometry, will the  gas  continue to flow-in  along preferred 
directions \citep[as does the dark matter, see e.g.][]{aubertetal04}, or does the hot phase erase any  anisotropy?
Will the above-mentioned smaller-scale non-linear dynamics eventually wash out any such trace?

Finally, note that the actual spin of the stellar disc at low redshift need not be trivially related to that of its larger-scale gravitational patch  \cite[see for instance][]{hahn10}, as a significant amount of angular momentum redistribution takes place in the circum-galactic medium \citep{kimmetal11} over  cosmic time.

\section*{Acknowledgments}
We thank J. Binney,  M. Haehnelt, S. Peirani, S. Prunet and T. Kimm for advice, and
 our collaborators of the  Horizon project ({\tt www.projet-horizon.fr}) for helping us produce the Horizon-4$\pi$ simulation. We also thank the anonymous referee for constructive
 criticism.
 The hydrodynamical simulation presented here was run on the DiRAC facility jointly funded
by STFC, the Large Facilities Capital Fund of BIS and the University
of Oxford. 
CP acknowledges support from a Leverhulme visiting professorship at the Physics department of the University of Oxford,  and thanks Merton College, Oxford for a visiting
fellowship. DP thanks the French Canada Research Fund and the University of Oxford.
JD and AS's research is supported by Adrian Beecroft, the Oxford Martin School and STFC.
Special thanks to T. Kimm for figure~\ref{fig:visual-nut},  F. Bouchet for allowing us to use the {\tt magique3} supercomputer during commissioning, and to S. Rouberol for making it possible.
We also thank D. Munro for freely distributing his Yorick programming language and {\tt opengl} interface (available at {\tt http://yorick.sourceforge.net/}).
\bibliographystyle{mn2e}
\bibliography{spin}

\bsp

\appendix
\section{Spin-tidal tensor correlations}
\label{sec:shear}

Let us present a complementary set of measurements: the correlations between the spin axis of dark halos, and  the orientation of the 
large-scale gravitational tidal tensor 
\mbox{$T_{ij} = \partial_{ij} \phi - \frac{1}{3} \Delta \phi \delta_{ij}$.}
To describe the orientation of  the tidal tensor,  we define
$\mathbf{e}_1$, $\mathbf{e}_2$ and $\mathbf{e}_3$ to be
the minor, intermediate and major eigen-directions of $T_{ij}$
according to the sorted eigenvalues 
$\lambda_{1}\le\lambda_{2}\le\lambda_{3}$ 
of the Hessian of the gravitational potential,
$\partial_{ij} \phi$,
(with which the tidal tensor shares 
the eigendirections). 

\begin{figure}
\includegraphics[width=1.\columnwidth]{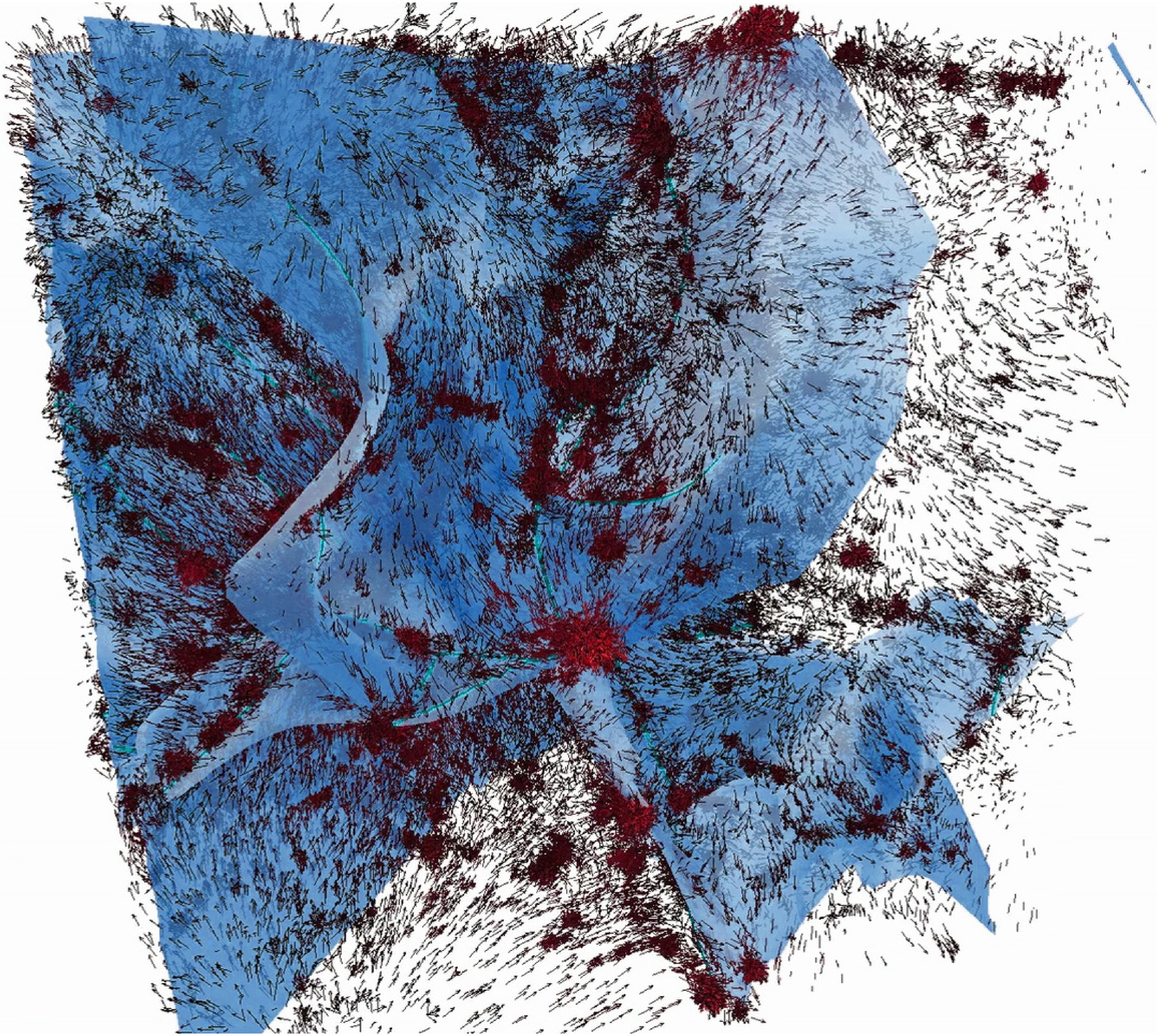}
\caption{The walls of the potential (in blue) and its filaments (in {\sl cyan}), together with the velocities of dark matter particles (in {\sl red})
of a 20 Mpc$/h$ $512^3$ dark matter particle $\Lambda$CDM simulation with WMAP-1 cosmogony. At redshift zero,
most dark matter halos sit in these walls, while the velocity field empties the voids and flows within those walls. 
This divergent flow is best seen in the top (hence bottom) left void (see also the animations online).  
\label{fig:walls} 
}
\end{figure}
Besides works based on the correlations between spin orientation and the cosmic web described in the main text (Section \ref{sec:intro}), 
the only numerical study to date looking at the alignment between halo spin and tidal tensor was done by \citet{b15} who predicted its orthogonality with the major principal axis but also found that galactic 
spins must have lost their initial alignment with the tidal tensor predicted by TTT. Direct observations of the alignment between the spin and the tidal tensor eigenvectors have also been carried out: the first attempt by \citet{lee02} studied the correlations between the disk orientation of the galaxies from the Tully catalog and the shear reconstructed from The Point Source Catalog Redshift survey and rejected the hypothesis of randomness at a 99.98 \% confidence level. More recently, \citet{b13} detected some correlations between the spin and the intermediate eigenvector of the tidal tensor and found that galactic spins were also preferentially perpendicular to the major
principal axis 
but this signal remains weak.
To overcome this lack of a clear numerical detection, 
the use of the 43 million halo sample of the Horizon 4$\pi$ simulation 
presents a tremendous advantage, as it allows us to very robustly calculate the correlations between halo spin orientations and the local tidal tensor.
We show in this Appendix that these measurements are not only consistent with the spin-filament correlations,
 but actually lend additional support to the interpretation of this paper in terms of large-scale dynamics along walls and filaments.
 
At the onset of non-linearity, the gravitational potential
tracks the velocity potential of the matter flow. Thus, the signs of
$\lambda_i$'s determine whether the flow in the corresponding direction
compresses ($\lambda_i > 0$) or rarifies ($\lambda_i < 0$)  matter.
At a given smoothing scale, this criterion can be used to partition
 space into
peak-like $0 < \lambda_{1}\le\lambda_{2}\le\lambda_{3}$, filament-like
 $\lambda_{1}<0$, $0< \lambda_{2}\le\lambda_{3}$, wall-like
$\lambda_{1}\le\lambda_{2} < 0$,   
$0< \lambda_{3}$ or void-like
$\lambda_{1}\le\lambda_{2}\le\lambda_{3} \le 0$, regions 
\citep{Pogosyan1998}. 
From this point of view,  in the peak regions
 matter compression is strongest along $\mathbf{e}_3$ and weakest
along  $\mathbf{e}_1$.  In the filamentary regions, 
$\mathbf{e}_1$ gives the direction of the filament, while the walls
are collapsing along $\mathbf{e}_3$ and extend, locally, in the 
plane, spanned by $\mathbf{e}_1$ and $\mathbf{e}_2$. 

At this stage, it is important to note that  the tidal field probes larger-scale structures than the filamentary structure studied in Section \ref{sec:discussion} as the gravitational potential is a smoother version of the density field (through  Poisson's equation). 
In other words, the skeleton of the potential (which locally corresponds to the eigen-directions of the tidal tensor) traces the cosmic structures (walls, filaments,\dots) on scales much larger than the skeleton of the density. Thus, 
 if we turn to a formulation in terms of the skeleton of the potential, the filaments described in Section \ref{sec:discussion} are embedded in the large-scale walls of the potential field
(as illustrated in fig.~\ref{fig:walls}), therefore protogalaxy formation begins with the first collapse (namely the collapse of $\mathbf{e}_{3}$) leading to the formation of the large-scale walls because it corresponds on smaller scale to the time when the filaments (of the density field) form by winding.
 
Note that a stricter definition of a filament in the local theory of the
skeleton  \citep{b6}
as a ridge in the density
profile, associates its local direction with the minor eigen-direction
of the Hessian of the {\sl density},
(here, to match enumeration, taken with a negative sign) 
$\mathbf{e}^{\rho}_1$.
Thus,
as the potential is two derivatives away from the density, 
the excess probability of alignment between the halo spin and $\mathbf{e}_1$  should be quite similar to that  of the alignment between the halo spin and the filament's direction.

\subsection{Alignment between spin and tidal eigen-directions}

\begin{figure}
\includegraphics[width=0.875\columnwidth]{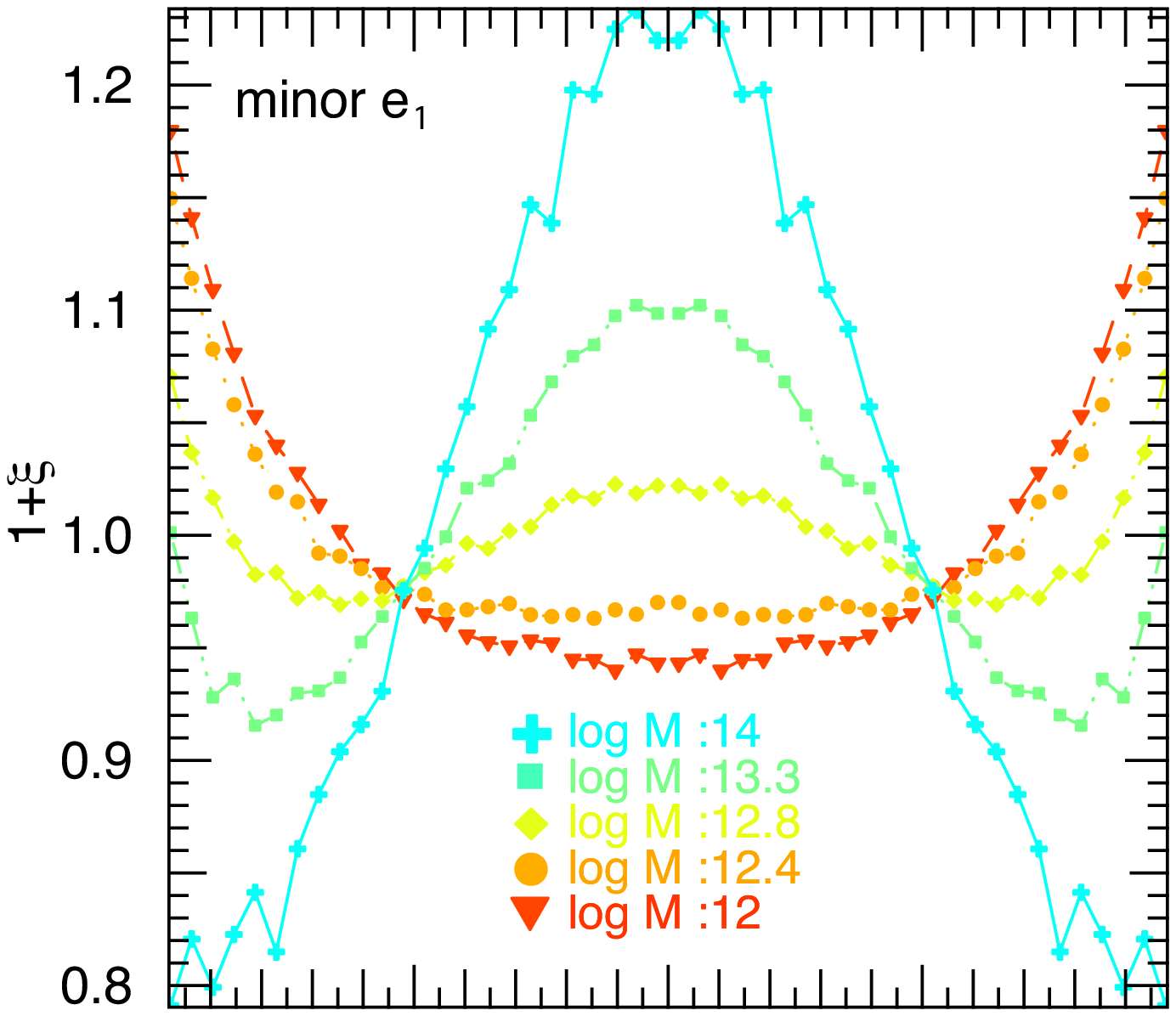}
\includegraphics[width=0.875\columnwidth]{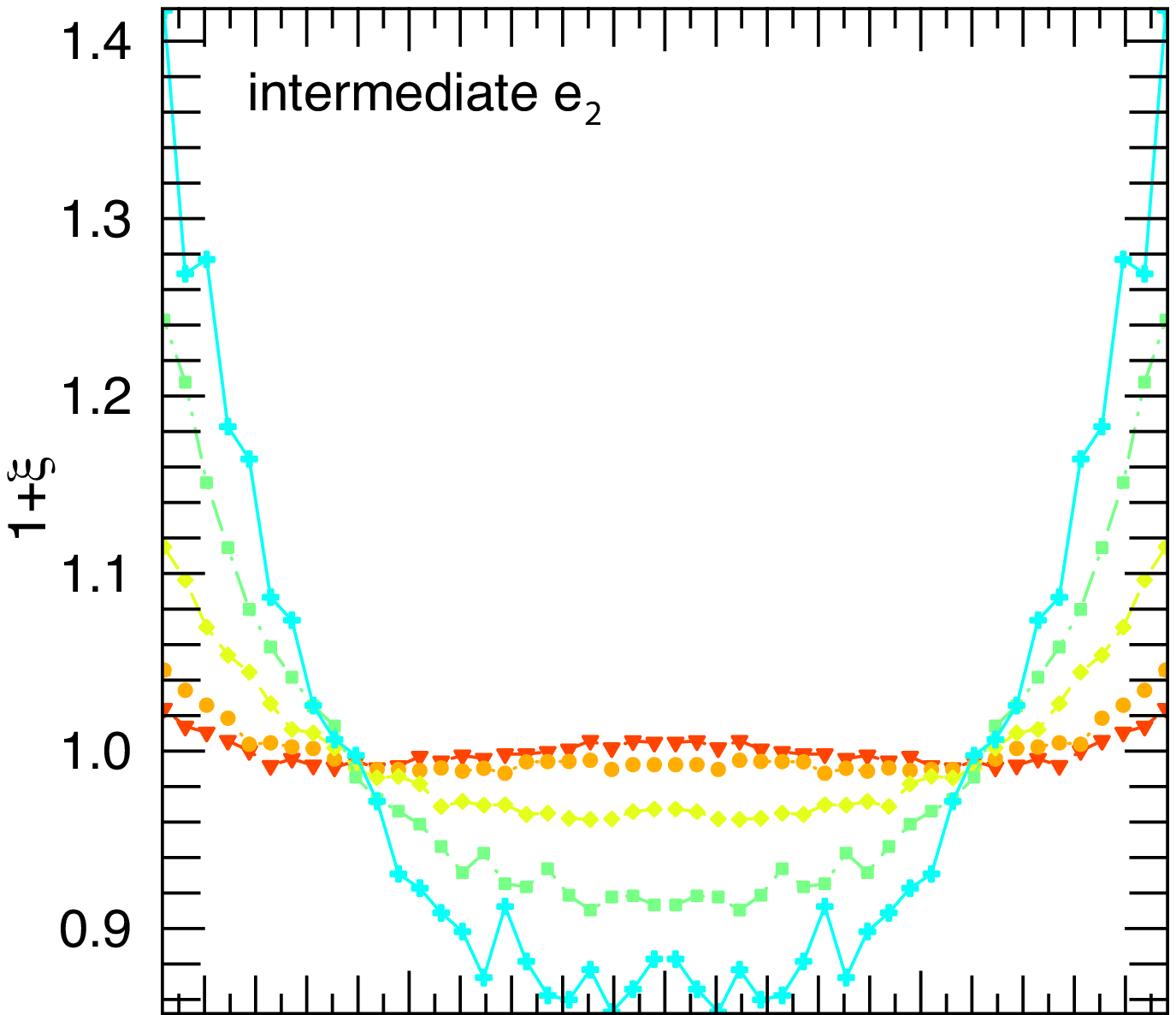}
\includegraphics[width=0.875\columnwidth]{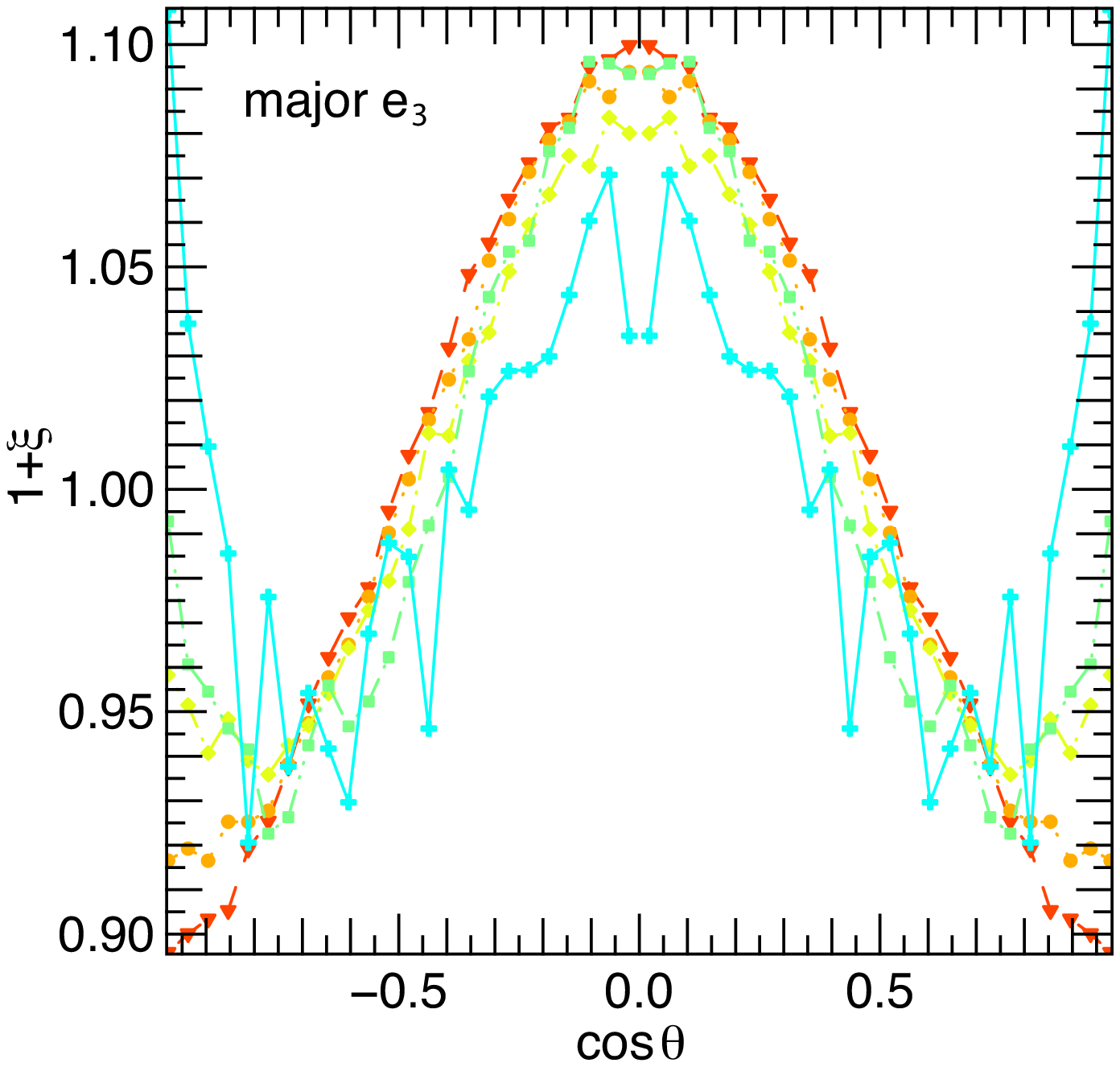}
\caption{Excess probability of alignment between the spin and the minor/intermediate/major axis ({\sl from top to bottom}) 
of the tidal tensor in the Horizon 4$\pi$ simulation.
Different mass bins are color-coded from $  10^{12}$({\sl  red}) to $10^{14}$ $M_\odot$ ({\sl blue}).
A transition is detected: spin of high-mass halos tend to be aligned with the
intermediate principal axis
({\sl middle panel}), whereas low-mass halos spins are
more likely to point along the minor axis
({\sl top panel}). 
\label{fig:shear} 
}
\end{figure}

In order to compute the excess probability of alignment between the spin of dark matter halos and the eigenvectors of the tidal tensor, the density field is again smoothed over 5 $h^{-1}$Mpc (for z=0), Poisson's equation is solved via Fast Fourier Transform 
in smaller overlapping boxes (boundary effects are found to be insignificant inside the boxes), the Hessian matrix of the potential is computed using a finite difference scheme ans
this matrix is finally interpolated to halo positions.
We then measure the angle between the angular momentum vector of the halo and each eigen-direction of the tidal tensor
and compute the histogram of the absolute value of the cosine of this angle; after normalization, it gives $1+\xi$, the excess probability of alignment between the  spin and the tidal tensor eigen-directions. As
 in Section \ref{sec:spin-fil}, the data is then split by halo mass.
 
Fig. \ref{fig:shear} displays these excess probabilities 
in three panels corresponding to the orientation of the spin w.r.t $\mathbf{e}_1$, $\mathbf{e}_2$ and $\mathbf{e}_3$ (top, middle and bottom panel,
respectively), at \mbox{$z=0$}.
 Halos of all masses have spins that preferentially avoid  
 the direction of the strongest large-scale compression, $\mathbf{e}_3$, 
 at 10\% excess probability ({\sl bottom panel}). The 
 exception is the very highest mass bin, which shows additional 
 halo population with
 spins aligned with $\mathbf{e}_3$.
 
For the high-mass, $M > M_{0}^t \approx 8(\pm 2)\cdot 10^{12}M_{\odot}$, halos
we detect a strong trend for the spin to be aligned with $\mathbf{e}_2$,
the intermediate principal axis of the tidal tensor with an excess probability
of up to 40\% ({\sl blue and green lines, middle panel}),
and to be perpendicular to the minor,
$\mathbf{e}_1$, principal axis with an excess probability of up to 20\%
({\sl blue and green lines, top panel}), a result in agreement with \citet{b13,b15}. 

Spins of the lower-mass, $M < M_{0}^t $,  halos tend, in contrast,
to align with $\mathbf{e}_1$ 
({\sl red and orange lines, top panel})
thus, preferring
the direction of the filamentary structures  $\mathbf{e}_{1}$,
with an excess probability of up to 15\%
and in a weaker way to align with $\mathbf{e}_{2}$ (with an excess probability of 5\%, see red and orange lines in the middle panel). 

These results are in exact agreement with our findings using the 
skeleton  that the spins of sufficiently large halos 
prefer to be perpendicular to the filament's direction while small halos
show a positive correlation for a spin orientation along the filaments. Indeed, the bottom panel of fig. \ref{fig:shear} is almost identical to the spin-filament correlation found in fig. \ref{fig:mass}.
For the tidal tensor, the transition occurs at a somewhat higher mass, 
$M_{0}^t \approx 8\cdot 10^{12}M_{\odot}$, than for the skeleton probe.
This is not surprising, 
because of the effectively larger scales probed by the tidal tensor and 
the observation (see Section \ref{sec:spin-fil} and Appendix~\ref{sec:smoothing})  that the critical mass increases with the smoothing length used to define the large-scale structure.

The redshift-dependence of the transition mass was also investigated (as described in the main text, see Section~\ref{sec:redshift}) and is found to be:
\[
 M^t_{\rm crit}\approx M^t_{0} (1+z)^{-\gamma_t},\,  \, \gamma_t=3 \pm 0.3 \textrm{ , }\,\, M^t_0\simeq 8(\pm 2) \cdot 10^{12} M_{\odot}. \nonumber
\]

\subsection{Consistency with LSS cosmic flows}

Let us  describe this spin acquisition in the framework of large-scale dynamics and the ellipsoidal collapse model \citep[among others]{lynden-bell64,zeldovich70,icke73,white79,peebles80,lemson93,bond96,sheth01,desjacques08}.
Low-mass halos form by accretion at high redshift. At this time, sheets and filaments
are forming by the successive collapses of (resp.) $\mathbf{e}_3$ and $\mathbf{e}_2$. When $\mathbf{e}_3$ is collapsing, low-mass halos form by accreting particles whose motion is along the direction $\mathbf{e}_3$. This process induces a spin perpendicular to this direction, i.e. in the plane $(\mathbf{e}_1,\mathbf{e}_2)$ which is the plane of the large-scale wall in which they are located. Then, $\mathbf{e}_2$ begins to collapse and other low-mass halos form from objects moving in the plane $(\mathbf{e}_2,\mathbf{e}_3)$ acquiring therefore a spin perpendicular to this plane i.e. aligned with $\mathbf{e}_1$ (which is the direction of the forming large-scale filament). 
The first generation of dark matter halos (of typically low mass) is now formed; as mentioned in Section \ref{sec:discussion}, for low mass halos at redshift zero,
 the spin will not change much as they have already acquired most of their mass. Fig. \ref{fig:MT2} (left panel) provides a clear illustration for this: a low-mass halo forms and acquires a spin aligned with $\mathbf{e}_1$ as expected. 
 It has been pointed out above that halo spin orientation must be either aligned with $\mathbf{e}_1$ (correlated to the filament's direction), or in the plane $(\mathbf{e}_1,\mathbf{e}_2)$ depending on the time 
at which they form. This is in good agreement with the top and middle panels of fig. \ref{fig:shear} which shows an excess probability for their spin to be aligned with $\mathbf{e}_1$ (or with the filaments) and in a weaker way with $\mathbf{e}_2$ (see the red and orange lines which represent the low-mass halos at redshift zero i.e. the halos with a mass between $3\cdot 10^{11}$ and $3\cdot 10^{12} M_{\odot}$).

Later (as described  in Section \ref{sec:discussion} in terms of flows along the filaments), $\mathbf{e}_1$  collapses and the halos located in the filaments of the potential stream along this direction. Most of the more massive halos then form by mergers in this flow and therefore acquire a spin which combines the spin of their progenitors and the orbital spin provided by the merger. The orbital spin must be in the plane perpendicular to $\mathbf{e}_{1}$ i.e. $(\mathbf{e}_2,\mathbf{e}_3)$ because the progenitors move along the filaments before merging, whereas the spin of their  (less massive) progenitors is in the plane $(\mathbf{e}_1,\mathbf{e}_2)$. The resulting angular momentum is therefore a superposition of these various spins, statistically more likely to be aligned with $\mathbf{e}_2$, which is  what was shown on fig. \ref{fig:MT1} ({\sl right panel}) and which is also in good agreement with the middle panel of fig.  \ref{fig:shear}, where the blue and green lines representing halos above  $2\cdot 10^{13}$ M$_\odot$ reveal a strong trend for these high-mass halos to be aligned with $\mathbf{e}_{2}$.
For these massive dark matter halos, the competition between the orbital spin and the intrinsic spin during the merger process was already pointed out
  in Section \ref{sec:discussion}. 
  The excess  probability for their spin to be aligned with $\mathbf{e}_2$ suggests that neither one nor the other dominates.
Nevertheless, very massive halos (with masses above $10^{14} $$M_{\odot}$) represented with a blue line in Fig \ref{fig:shear} seem to have their spin less perpendicular to $\mathbf{e}_3$: actually we can observe two modes, one perpendicular and one aligned with this direction, which can be understood if they are  the result of a further generation of mergers  whose intrinsic spins were already perpendicular to the filament.

Let us emphasize that this explanation and that which was presented in Section \ref{sec:discussion} are consistent.
Indeed, in Section \ref{sec:discussion} our claim is that winding of the walls is responsible for the direction of the spin of low-mass halos.
Meanwhile, in this Appendix, the focus is on the first collapse along  $\mathbf{e}_3$   i.e. on the formation of walls. However, as we was pointed out in the introduction 
of this Appendix, the tidal field probes {\em larger} scale structures than the filaments of the density field studied in Section \ref{sec:discussion}. 
Taking into account this difference, it turns out that the two analyses are complementary (describing the large-scale dynamics on different scales). 
It also helps reconciling the findings of \citet{b15,b13} which rely on the tidal tensor with those of \citet{b24,b10,b14,b16}  which involve the density field filamentary structure.

\section{ Circum-galactic medium spin } 
\label{sec:ap-hydro}

\begin{figure}
   \includegraphics[width=1.\columnwidth]{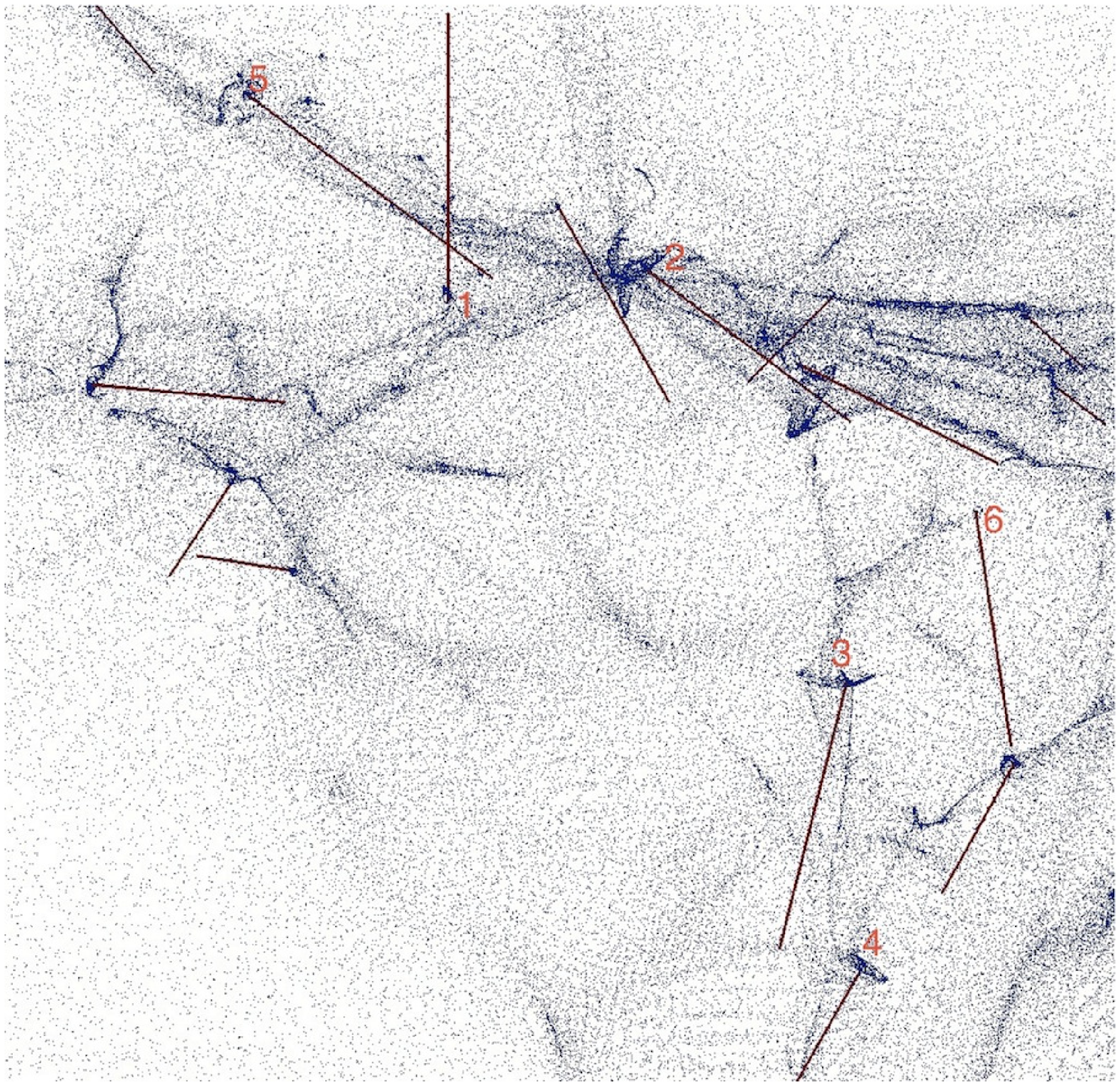}  	
   \vskip 0.1cm
   \includegraphics[width=1.\columnwidth]{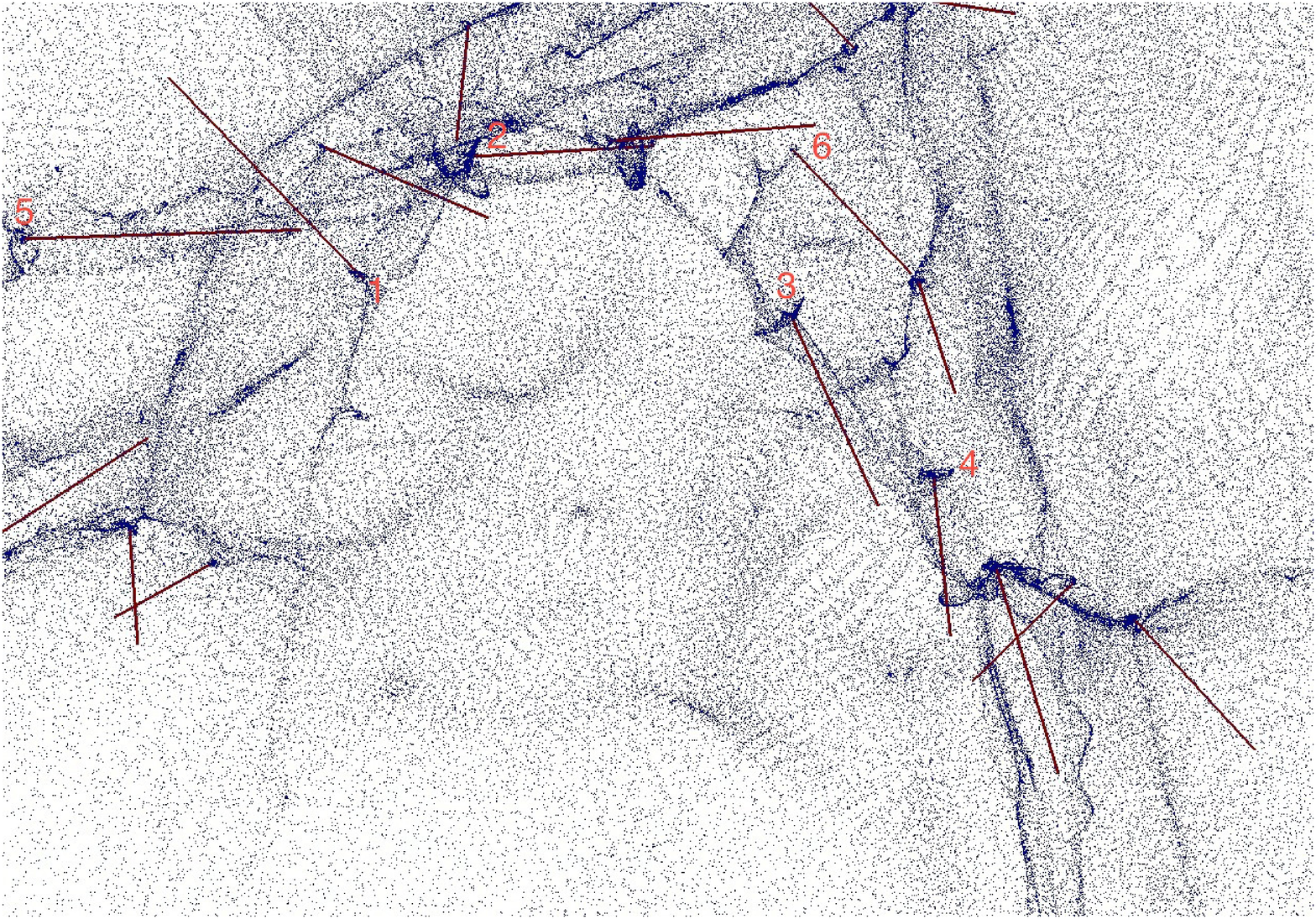}  	
\caption{Distribution of hydrodynamical tracer particles ({\sl blue}) at redshift 9, along two different viewing angles. The spin direction of the circum-galactic medium between $0.1 R_{vir}$ and $0.25 R_{vir}$ of some protogalaxies is plotted in  red. On average the spins are more likely to be aligned with the filaments (as seen with protogalaxies 2,3,4 and 5 for example).   
}
\label{fig:spin-yohan}
\end{figure}

Let us first describe briefly the {\sl hydrodynamical} simulation that we will use to assess the alignment of the gas component surrounding the galaxies with the large-scale filamentary structure. This simulation is described in more details in~\cite{Dubois2011} and corresponds to their SHhr run. Let us recall here its basic properties. The SHhr simulation follows the formation of a massive ($M_{\rm vir}=5\cdot 10^{11}\, \rm M_\odot$ at $z=6$) halo within a re-simulated region in a 100 Mpc$/h$ box size, its dark matter mass resolution is $1.3\cdot 10^6 \, \rm M_\odot$, and minimum cell size is 17 pc. The cosmology employed in that run is slightly different from the parameters of the Horizon 4$\pi$ simulation, and is compatible with the WMAP7 cosmology \citep{komatsuetal11}: $\Omega_{\rm m}=0.27 $, $\Omega_{\Lambda}=0.73$, $\Omega_b=0.045$, $H_0=70 $ km$\cdot s^{-1} \cdot $Mpc$^{-1}$, $n=0.961$, and $\sigma _8=0.8$.
The gas is allowed to cool down radiatively down to $T_0=100$ K, assuming an initial metal enrichment of $10^{-3}\, \rm Z_\odot$ \citep{sutherland&dopita93}. A UV background heating source term is added to the gas energy equation following \cite{haardt&madau96} with reionization taking place at $z_{\rm reion}=8.5$. Star formation is allowed in gas density regions above $n_0=50 \, \rm H\cdot cm^{-3}$ using a Poisson random process \citep{rasera&teyssier06, dubois&teyssier08winds} that reproduces the Schmidt-Kennicutt law $\dot \rho_*=\epsilon_* \rho /t_{\rm ff}$, where $\dot \rho_*$ is the star formation rate density, $\epsilon_*=0.01$ the star formation efficiency, and $t_{\rm ff}$ the local free fall time of the gas with local density $\rho$. No feedback from supernovae or Active Galactic Nuclei is accounted for. Tracer particles that passively follow the motion of the gas are scattered in the initial conditions and allow us to trace back the Lagrangian trajectories of gas elements that end up in collapsed structures.

Fig.~\ref{fig:spin-yohan} displays  two large-scale views of this  hydrodynamical simulation and its tracer particles at redshift 9 shown in fig.~\ref{fig:visual-tracer}.
 Structures and substructures in dark matter are detected with the Most-massive Sub-node Method \citep{b37} and only masses above $5 \cdot 10^{8} M_{\odot}$ 
are selected. The spin of the circum-galactic medium (accounting for non-star forming gas only with gas density below $n_{\rm H}<50 \, \rm H.cm^{-3}$) between $0.1 R_{vir}$ and $0.25 R_ {vir}$ is then computed and its orientation is represented with dark red segments. 
This  somewhat ad-hoc criterion  used to define the spin  reflects our focus on  the angular momentum of the secondary infall gas, which has just been or is being accreted;  our measurements correspond to an average of the spin within this subregion of the dark matter halo.
This figure  shows a good alignment of the spins with the circum-galactic polar filaments (i.e. the filament which visually flows along the polar axis of the galaxy, 
in particular for the clumps 2,3,4,5).
This is consistent  with  visual inspection (fig.~\ref{fig:visual-tracer}, but best carried online),
 and with the prediction of fig.~\ref{fig:mass}, since the critical mass at that redshift is $6\cdot 10^{10}M_{\odot}$. 
The spin of a  couple of  low-mass clumps in that field (noticeably clump 1 and 6), are in fact poorly estimated automatically, as the dark matter clump center can be offset at 
that redshift relative to that of the circum-galactic disc.

\section{Robustness of  spin-filament correlation} 
\label{sec:checks}

Let us assess  how robust the  spin orientation - filament correlation found in Section~\ref{sec:spin-fil} is by carrying a few  consistency tests. 

\begin{figure}
 \includegraphics[width=0.95\columnwidth]{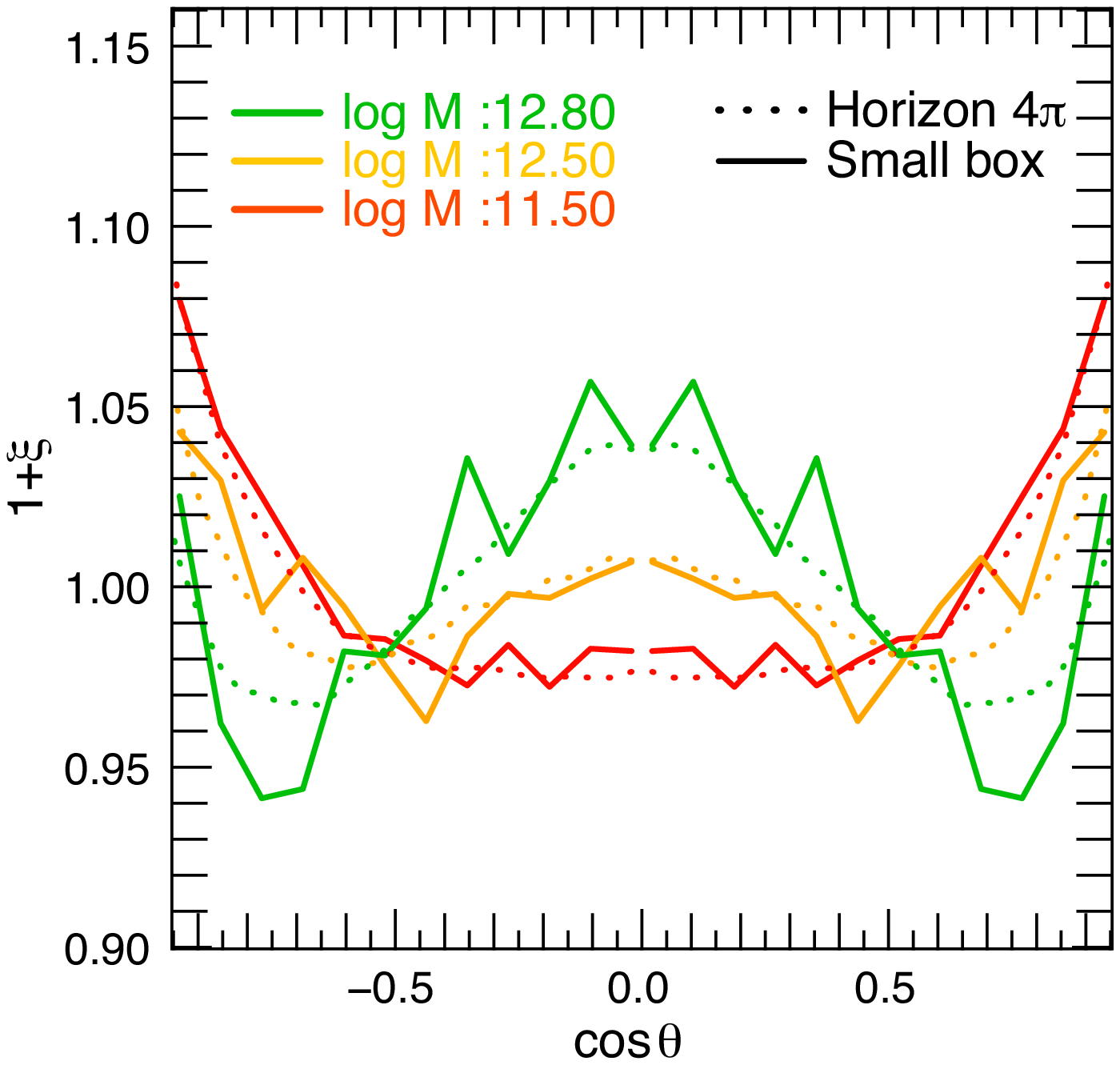}
 \caption{Excess probability of  spin-filament alignment in various simulations.
 Measurements in the Horizon 4$\pi$ simulation are plotted with dotted lines and measurements in the smaller simulations are plotted with solid lines.
 Different colors correspond to different halo mass bins from $3\cdot10^{11}$({\sl red}) to $6\cdot10^{12}$ $M_\odot$ ({\sl green}). 
 The signal in the Horizon 4$\pi$ simulation is statistically consistent  with  that of the smaller simulations.
 }
 \label{fig:checksmallsimus}
\end{figure}

In order to check the effect of mass resolution (the spin of very low-mass halos is poorly defined for instance, as too few particles are involved in its measurement),
 the same measurement are carried out in a smaller simulation ($256^3$ particles in a $50$ $h^{-1}$Mpc periodic box with the same cosmology) for which same physical masses are represented by higher numbers of particles: mass bins from $3\cdot10^{11}$({\sl red}) to $6\cdot10^{12}$ $M_\odot$ ({\sl green}) correspond to 30-1400 particles in Horizon 4$\pi$ (corresponding to a lower threshold for the FoF detection) since the mass per particle is $7\cdot10^{9}$ $M_\odot$; in contrast,  350 to 15000 particles are found in halos of the same mass for the small simulation since the mass per particle is $6\cdot10^{8}$ $M_{\odot}$. The detection of the the same phase transition (see fig. ~\ref{fig:checksmallsimus}) occurring at the same halo mass demonstrates that the signal is not induced by limited mass resolution. This result is also consistent with the findings of  \citet{b10,b14}.

Another simple check involves   varying the procedure   by choosing for each halo the closest segments of the skeleton (instead of for each segment, the closest halos). The signal we get is very similar to fig. \ref{fig:mass} (low-mass halos tend to be parallel to the filaments with an excess probability of 15\% and high-mass halos perpendicular with an excess probability reaching 20\% for the more massive bin which is even stronger than in fig. \ref{fig:mass}), suggesting that the measured correlations are independent from the detailed procedure implemented to identify neighbours.

Halos at the nodes of the skeleton cannot have a well defined closest-segment direction, as more than one skeleton segment typically qualifies, and could therefore bias our measurements. So as to quantify this effect, the same algorithm is implemented but with a new criterion: halos closer than a certain distance to the nodes are not considered. An even stronger signal is detected, which leads us to conclude that nodes introduce extra noise and are not the cause for the observed signal.

One might also think this result could depend on a density threshold for the underlying filament. Hence, the same data in low, middle and high-density filaments were plotted: the excess probability of spin-filament  alignment  is found to be the same whatever the density inside the filaments is.
%
All these  tests demonstrate the overall robustness of the mass-dependent transition of the relative alignment.

\section{Characteristic masses}
\label{sec:masses}

In the main text the  density field was smoothed over  a scale of 5 Mpc$/h$ corresponding to a mass of $1.9 \times 10^{14} M_\odot$.
The transition mass found in Sec.~\ref{sec:spin-fil} is therefore defined relatively to this mass.
In the context of hierarchical clustering, as long as the field is smoothed on scales where filaments are still well defined,  
one can anticipate some scaling of this transition mass with smoothing. This transition mass should
 reflect the connection between the geometry of the larger scale flow and the  mass scale corresponding to galaxies forming and drifting on this cosmic web.

\subsection{Smoothing-dependence of the critical mass}
\label{sec:smoothing}
\begin{figure}
  \includegraphics[width=\columnwidth]{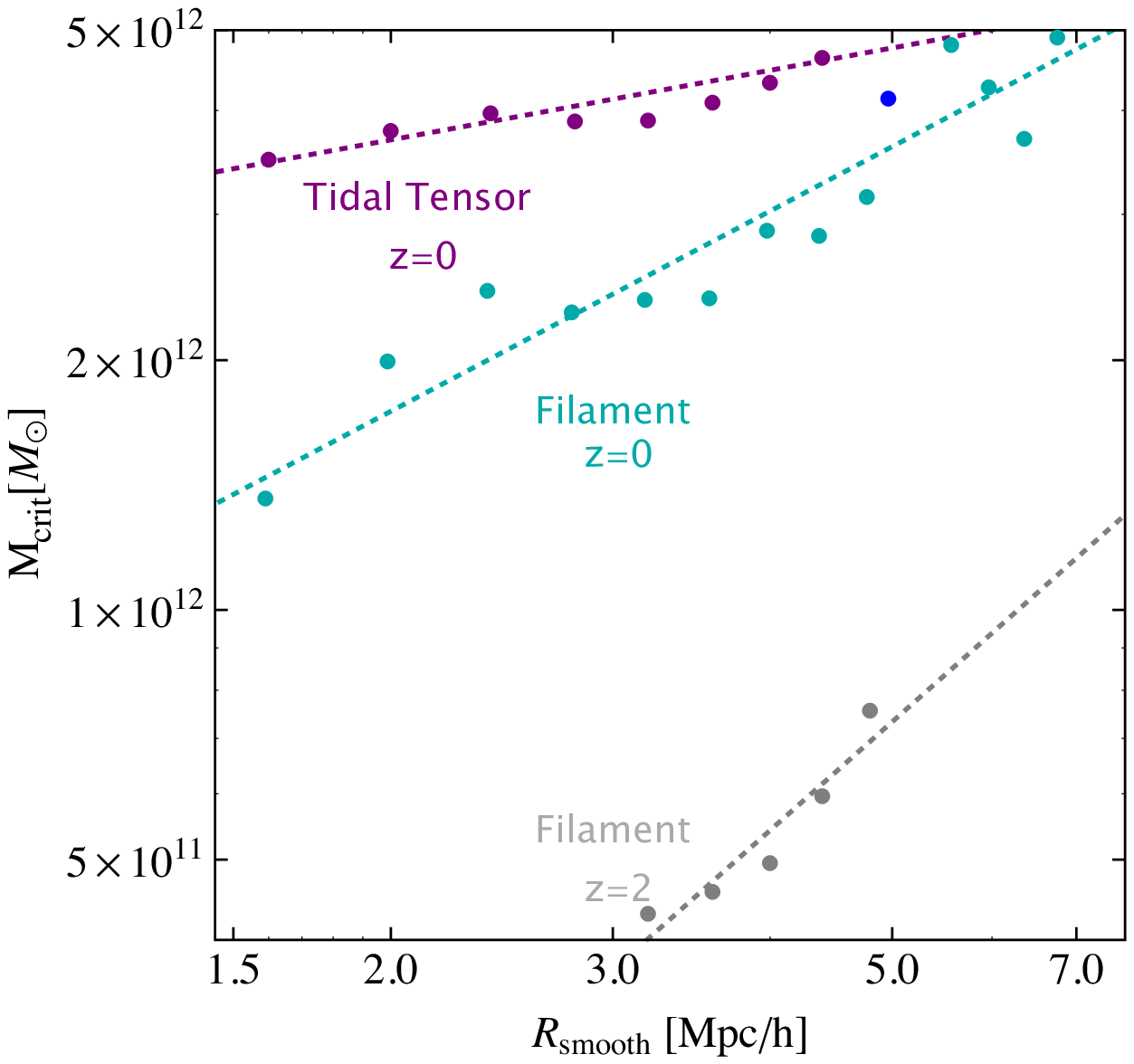}
  \vskip -1cm
 \caption{Evolution of the critical mass as a function of the smoothing length for the filaments at redshift 0 ({\sl cyan}) and 2 ({\sl grey}) and the tidal tensor at redshift 0 ({\sl purple}). The critical mass increases with the smoothing length with a redshift-dependent slope. The blue dot was measured in the 4$\pi$ simulation for the filaments.}
 \label{fig:smoothing}
 \label{fig:ms_vs_R}
\end{figure}

Fig. \ref{fig:smoothing} displays the evolution of the critical mass with the smoothing length. It is found that  $M_{\rm crit}^{s}( R )$ and $M_{\rm crit}^{t}( R )$ can be well fitted by power-laws, namely:
\begin{equation}
M_{0}( R )\simeq M_{0}(R_{0})\left(\frac{R}{R_{0}}\right)^{\alpha}\,,
\end{equation}
where $M^{s}_{0}(R_{0})\simeq 3.6(\pm1) \cdot 10^{12}$, $\alpha^{s}\simeq 0.8 \pm 0.1$, $M^{t}_{0}(R_{0})\simeq 4.8 (\pm 1)\cdot 10^{12}$ and $\alpha^{t}\simeq 0.28 \pm 0.04$.
This scaling is   not inconsistent with the discussion of the origin of the alignment given in Section 3 in  as much as  
a smoothing length defines a set of filaments and therefore picks out 
a halo mass scale corresponding to the halos which are flowing/merging along these filaments. Note that in practice the dependence on smoothing is actually rather weak
(expressed in terms of mass, we have $M_{\rm crit} \propto M_{\rm smooth}^{0.27}$ for filaments
defined via the skeleton and $M_{\rm crit} \propto M_{\rm smooth}^{0.09}$ 
if structure anisotropy is described via the shear tensor).  
Fig. \ref{fig:smoothing} also shows that $\alpha$ depends on redshift. Note that as expected, the critical mass for the tidal tensor matches that of the filaments smoothed on a larger scale ($\simeq 7.5 $ Mpc$/h$ instead of 5). This confirms the idea that the potential is close to a smoother version of the density field.

\subsection{Non-linear mass  evolution}
\label{sec:masstransition}
It is of interest to compare the transition masses, 
$M^s_{\rm crit}(z)$ and $M^t_{\rm crit}(z)$ to the mass scale that
tracks the development of non-linearity in structure formation.
The variance of the density field smoothed on scale $R$ obeys
\begin{equation}
\sigma^2(R,z)= D(z)^2 \int_0^\infty \! \! P(k) W^2(k R)  {\rm d}^3 k,\,
\end{equation}
with $P(k)$ the powerspectrum, and the top-hat filter
defined by $W^{2}(x)= {9}\left({\sin x}/{x}-\cos x\right)^{2}/{x^{2}}$.
The growth factor, $D(z)$ is given by
\begin{displaymath}
D(z) =  \frac{5}{2} \Omega_m H_0^2  H(z)
\int_z^\infty \frac{ (1+z) {\rm d} z}{ H(z)^3},\,
\end{displaymath}
 with $ H(z)=H_0 \sqrt{\Omega_m (1+z)^3 + \Omega_\Lambda}$.
 Here $H_0$, $\Omega_m$ and $\Omega_\Lambda$ are the Hubble constant, the dark matter  and the dark energy density parameters 
 at $z=0$ respectively.

Fixing the level of (non-)linearity by the condition $\sigma(R(z),z) = const$
implicitly defines the redshift evolution of the smoothing scale $R(z)$ (expressed in comoving
$\rm Mpc$) that maintains
this level of (non-)linearity. This, in turn, corresponds to the mass scale
\begin{equation}
\label{eq:mass}
 M_{\rm NL} (z)\equiv \frac 4 3 \pi  {\bar\rho} R(z)^3\,,
 \end{equation}
where ${\bar \rho}$ is the present-day average density of matter in the Universe.

For the matter dominated CDM Universe with a 
scale-free power spectrum $P(k) \propto k^n$, 
$M_{\rm NL}(z) \propto (1+z)^{-{6}/{(n+3)}}$.
In the Universe with realistic parameters,
the redshift dependence of $M_{\rm NL}(z)$ is not a power law,
both due to the influence of the $\Lambda$-term that slows the growth of
the structure at low redshifts, and steepening of the spectrum as one moves to
smaller scale at high redshifts.

\label{lastpage}

\end{document}